\documentclass[11pt,reqno]{amsart}
\usepackage[T1]{fontenc}
\usepackage{lmodern}
\usepackage{mathtools,amssymb,amsthm,mathrsfs}
\usepackage[margin=1.05in]{geometry}
\usepackage{microtype}
\usepackage[shortlabels]{enumitem}
\usepackage{xcolor}
\usepackage[colorlinks=true,linkcolor=blue!45!black,
  citecolor=blue!45!black,urlcolor=blue!45!black]{hyperref}
\newtheorem{theorem}{Theorem}[section]
\newtheorem{proposition}[theorem]{Proposition}
\newtheorem{lemma}[theorem]{Lemma}
\newtheorem{corollary}[theorem]{Corollary}
\theoremstyle{definition}
\newtheorem{definition}[theorem]{Definition}
\theoremstyle{remark}
\newtheorem{remark}[theorem]{Remark}
\numberwithin{equation}{section}
\newcommand{\C}{\mathbb C}
\newcommand{\R}{\mathbb R}

\newcommand{\Z}{\mathbb Z}
\newcommand{\T}{\mathbb T}
\newcommand{\dd}{\,d}
\usepackage{xurl}
\newcommand{\eps}{\varepsilon}
\newcommand{\leanname}[1]{\nolinkurl{#1}}

\DeclareMathOperator{\sgn}{sgn}

\DeclareMathOperator{\tr}{tr}
\setlist[enumerate]{label=\textup{(\roman*)},leftmargin=*,itemsep=3pt,topsep=5pt}
\setlist[itemize]{leftmargin=*,itemsep=3pt,topsep=5pt}
\allowdisplaybreaks[1]
\title[Ten Dry Martinis, Five on the Rim]{Persistence of Unitary Almost Mathieu Operator Gaps:\\ Ten Dry Martinis, Five on the Rim}
\author{Dan S. Borgnia}
\address{Department of Physics,
University of California, Berkeley,
California 94720, USA}
\email{dborgnia@berkeley.edu}

\author{Matthew Faust}
\address{Department of Mathematics,
Michigan State University,
East Lansing, Michigan 48824, USA;
Simons Institute for the Theory of Computing,
University of California, Berkeley,
California 94720, USA}
\email{mfaust@msu.edu}

\author{Long Li}
\address{Department of Mathematics,
Texas A\&M University,
College Station, Texas 77843, USA}
\email{longli@tamu.edu}
\date{}
\begin{document}
\begin{abstract}
We determine exactly which gap labels are realized by open spectral gaps for the unitary almost Mathieu operator at every irrational frequency and all coupling constants ($0<\lambda_1,\lambda_2\leq1$). When both couplings are strictly less than one, every permitted gap label is realized by an open gap, including in the critical case ($\lambda_1=\lambda_2$). When at least one coupling equals one, both lifts of every odd winding label are realized by open gaps, whereas all even winding labels are collapsed. The proof combines quantitative almost reducibility and Aubry duality with a monotonicity principle for gap heights.
\end{abstract}
\maketitle

\smallskip\noindent\textbf{\href{https://mattfaust.github.io/Dry10/uamo\_LEAN.zip}{Lean verification} convention.}
Following \cite{BorgniaFaust2026}:
$L^*$ denotes a checked version of the written argument, while $LE$
denotes a conclusion checked with an explicitly identified replacement
for a cited input or proof route. Appendix~\ref{app:lean-audit}
records the precise scope and declaration correspondence.
\medskip

\section{Introduction}~\label{Sec:Intro}
The famous ``Ten Martini Problem'' was initially posed by Kac who promised ten Martinis for the solution of the following problem \cite{Kac_martinis}: Consider the almost Mathieu operator (AMO)
\begin{align}\label{eq:AMO}
    (H_{\lambda,\alpha,\theta} \psi)(n)=\psi(n+1)+\psi(n-1)+2\lambda \cos2\pi(\theta+n\alpha)\psi(n),\quad\lambda>0,\:\alpha\in\mathbb{R}\setminus\mathbb{Q},\:\theta\in\mathbb{T}.
\end{align}
Kac asked ``are all the gaps [in the spectrum of the AMO] there?'' or, in other words, ``is the spectrum of the AMO a Cantor set?''. This problem was popularized by Simon \cite[Problem 1]{simonAlmostPeriodicSchrodinger1982} and  completely solved by Avila-Jitomirskaya \cite{AJ09}, building on considerable effort with partial progress along the way  \cite{bellissardCantorSpectrumAlmost1982,choiGaussPolynomialsRotation1990,Puig2004CMP}. For an in-depth overview over the development we refer the reader to the surveys \cite{marxDynamicsSpectralTheory2017,damanik2024two}.

The ``Dry Ten Martini Problem'' originally asks the natural follow up question \cite[Problem 2]{simonAlmostPeriodicSchrodinger1982}: {\it are all gaps in the spectrum of the AMO open?} It is by now fully understood to be true \cite{choiGaussPolynomialsRotation1990,Puig2004CMP,AJ2012JEMS,LiuYuan2015,AYZ24,BorgniaFaust2026}. One can of course ask the (Dry) Ten Martini Problem for any model of interest. Recent progress has been made for the extended Harper's model \cite{hanDryTenMartini2017}, Sturmian Hamiltonians \cite{raymond1997constructive,DGY2016,band2024dry,bandReviewWorkRaymond2024} and $C^2$- and $\cos$-type sampling functions \cite{WangZhang2017,FV21,GeWangXu2025} and perturbations of type-I operators \cite{GJY,GeJitomirskayaYou2026,Li2026,Ge2026,LiXuZhou2026}). For discontinuous monotone potentials, Cantor spectrum is established in \cite{KPS2025}, and all permitted gaps are open for generic sampling in the perturbative setting of \cite{KPS2026}. More general results in the regime of positive Lyapunov exponent can be found in \cite{GS2011} and the references therein.

We note that one can further abstract the question and ask how, for a family of random operators defined over an ergodic dynamical system, the base dynamics determines the topological structure of the almost sure spectrum \cite{DEF2024}. Clearly, the Dry Ten Martini Problem is a special case of this ``all gaps open'' problem. As for operators with quasiperiodic base dynamics such as the AMO, the set of possible gap labels is determined by the so-called Gap Labeling Theorem \cite{bellissardGapLabellingTheorems1992,bellissardKtheoryAlgebrasSolid1986,damanikGapLabellingDiscrete2023,johnsonRotationNumberAlmost1982a,GeronimoJohnson1996JDE}. The converse of this is the question whether a label predicted by the Gap Labeling Theorem labels an actual gap, i.e., a gap that is not collapsed or degenerate. This converse problem is in general very difficult since both the base dynamics and the sampling functions can affect the topological structure of the spectrum, compare \cite{ABD12}.

A crucial ingredient in the classical proof of the Dry Ten Martini Problem for the AMO is the \emph{Aubry-André} duality \cite{AJ2012JEMS,Puig2004CMP}, that is, a duality between different elements of the family $\{H_\lambda\}$ with respect to a twisted Fourier transform.
For more general quasiperiodic operators, many available results establish generic gap opening or generic Cantor spectrum \cite{eliassonFloquetSolutions1dimensional1992,PUIG_SIMO_2006,Puig2006a,ABD09,ABD12,DL24,KPS2026}. On the other hand, whenever one has this duality available for a concrete operator, one can expect the Dry Ten Martini problem to hold.

Here, we solve the Dry Ten Martini Problem for a model that goes beyond the class of quasiperiodic Schrödinger operators but  which  still exhibits the André-Aubry duality: the so-called \emph{unitary almost Mathieu operator} (UAMO) is a quantum walk that describes the motion of a single particle with two-dimensional internal degree of freedom on the integers in discrete time steps. The dynamics is described by the alternation of two types of operators: a parametrized shift and a ``coin'' operator that locally acts via a matrix-valued quasi-periodic function. This model was introduced and studied in \cite{CFO23}, see also \cite{fillmanSpectralCharacteristicsUnitary2017,CFGW19}. There, among other results concerning the spectral types in various parameter regimes, the Ten Martini Problem was solved on the critical line for all irrational frequencies \cite[Theorem~2.2(d)]{CFO23}; the corner $\lambda_1=\lambda_2=1$ was treated earlier in \cite{fillmanSpectralCharacteristicsUnitary2017}. 
This work advances the result by proving the ``dry'' version and showing that all gaps are open, up to a surprising exception when at least one of the coupling parameters has unit modulus, see Theorem~\ref{thm:classification}.

Let $\mathcal{H}=\ell^2(\Z)\otimes\C^2$ with basis $\{\delta_n^s:n\in\Z,s\in\{+,-\}\}$ where $\delta_n^s:=\delta_n\otimes e_s$ and $\{\delta_n\}_{n\in\Z}$ and $\{e_+,e_-\}$ are the standard bases of $\ell^2(\Z)$ and $\C^2$, respectively. 
The UAMO is a split-step quantum walk $W_{\lambda}=S_\lambda Q$ on $\mathcal{H}$ that is defined in terms of two operators: 
$S_\lambda$ is the conditional shift operator
\begin{align}
S_\lambda:\delta^\pm_{n}\mapsto\lambda\delta^\pm_{n\pm1}\pm\lambda'\delta^\mp_n,\qquad\lambda\in[0,1],\:\lambda'=\sqrt{1-\lambda^2},
\end{align}
and $Q$ is a ``coin'' operator that acts coordinate-wise via a $2\times2$ unitary $Q_n$. For the UAMO, these local coins are quasi-periodically distributed according to the rule
\begin{align} \label{eq:coindef}
Q_{\lambda,n}
=
Q_{\lambda,n,\alpha,\theta}
=
Q_{\lambda,2\pi(n\alpha + \theta)},
\quad
n \in \mathbb{Z},
\end{align}
where $Q_{\lambda,\theta}$ denotes a modified counterclockwise rotation through the angle $\theta$, i.e.,
\begin{align}\label{eq:rotation_coin}
Q_{\lambda,\theta}
=
\begin{bmatrix}
\lambda\cos(\theta)+i\lambda' & -\lambda\sin(\theta) \\
\lambda\sin(\theta) &  \lambda\cos(\theta)-i\lambda'
\end{bmatrix},
\qquad
\theta \in \mathbb{R}/(2\pi\mathbb{Z}),\:\lambda\in[0,1],
\end{align}
and we abbreviated $\lambda'=\sqrt{1-\lambda^2}$. Here the coin angle is measured in radians; the phase in \eqref{eq:coindef} belongs to $\mathbb T=\mathbb R/\mathbb Z$. 

With these two building blocks, the UAMO is defined as
\begin{align}\label{eq:model}
W_{\lambda_1,\lambda_2,\alpha,\theta}:=S_{\lambda_1}Q_{\lambda_2,\alpha,\theta},
\end{align} 
and we abbreviate as $W_{\lambda_1,\lambda_2}$ for fixed $\alpha$ and $\theta$. As discussed in detail in \cite[Section 3]{CFO23}, $\alpha$ plays the role of a magnetic field in an associated two-dimensional model, and $\theta$ plays the role of a Fourier parameter of the second lattice dimension. We shall nevertheless stick with the standard nomenclature in dynamical systems and call $\alpha$ the \emph{frequency} and $\theta$ the \emph{phase} and refer to $\lambda_1$ and $\lambda_2$ as \emph{coupling constants}, as they determine the ``strength'' of the shift and the coin, respectively.

\medskip
The UAMO displays a metal-insulator phase transition \cite{LinCedzichZhouXue2026}, with respect to the coupling constants for almost all $\alpha$ and $\theta$: for $\lambda_1>\lambda_2$ the spectrum of $W_{\lambda_1,\lambda_2}$ is purely absolutely continuous, for $0<\lambda_1=\lambda_2\leq1$ it is purely singular continuous while for $\lambda_1<\lambda_2$ it is pure point with exponentially decaying eigenfunctions. This follows from the spectral and cocycle classification in \cite[Theorems~2.2, 2.7 and 2.9]{CFO23}, with the cocycle terminology of Avila's global theory \cite{Avila2015Acta}.

We use this terminology in the remainder of the paper to refer to the respective parameter ranges. For example, we refer to the ``critical UAMO'' when we mean the UAMO with equal coupling constants $\lambda_1=\lambda_2$. Another important ingredient in these proofs is a unitary version of André-Aubry duality, which relates $W_{\lambda_1,\lambda_2}$ to its ``dual''
\begin{align}
    W_{\lambda_1,\lambda_2}^\sharp:=W_{\lambda_2,\lambda_1}^\top,
\end{align}
and which immediately implies that the spectra of $W_{\lambda_1,\lambda_2}$ and $W_{\lambda_2,\lambda_1}$ are the same.

For $0\le\lambda\le1$, write $\lambda'=\sqrt{1-\lambda^2}$. On
$\ell^2(\Z;\C^2)$ define
\begin{align}
 (S_\lambda u)^+_n=\lambda u^+_{n-1}-\lambda'u^-_n,
  \quad (S_\lambda u)^-_n=\lambda u^-_{n+1}+\lambda'u^+_n
\end{align}
and
\begin{align}
Q_\lambda(x)&=
\begin{pmatrix}
\lambda\cos2\pi x+i\lambda'&-\lambda\sin2\pi x\\
\lambda\sin2\pi x&\lambda\cos2\pi x-i\lambda'
\end{pmatrix}.
\end{align}
For irrational $\alpha$, its spectrum is independent of $\theta$; we denote it by $\Sigma_{\lambda_1,\lambda_2,\alpha}$. We use the  convention of \cite{CFO23} for the cocycle: $\lambda_1>\lambda_2$ is subcritical, $\lambda_1<\lambda_2$ is supercritical, and $\lambda_1=\lambda_2$ is
critical. 

The rotation number $\rho(t)$ is that of the degree-zero, two-step real
Szeg\H{o} cocycle $A_t$ defined in \eqref{eq:real-cocycle}, with $z=e^{it}$.
We use the integer $m$ as the \emph{winding number} of a gap:
its constant rotation-number value $\rho_*$ satisfies
\begin{align}\label{eq:labels}
 2\rho_*\equiv m\alpha\pmod1,\qquad m\in\Z.
\end{align}
For fixed $m$, this congruence has two solutions in $\R/\Z$,
differing by $1/2$. We do not introduce a separate index for them.
When following a particular gap, we retain its value $\rho_*$;
the assertion that \emph{all gaps with winding number $m$ are open}
means that each of these two values occurs on a nonempty open
spectral gap. A permitted value is called \emph{collapsed} if it
does not occur on such a gap. Irrationality of $\alpha$ makes $m$
unique for each permitted $\rho_*$. An overall orientation reversal
changes $m$ to $-m$ and does not change its parity.
At singular boundary couplings, gap values are defined by the
circular integrated density of states, its agreement with the
rotation-number convention is discussed in
Section~\ref{subsec:gap-continuation}.
The density-of-states and rotation-number conventions are those of
\cite{CFO23,CL26}.

\begin{theorem}[$LE$]\label{thm:interior}
Let $\alpha$ be irrational and let $0<\lambda_1,\lambda_2<1$ with
$\lambda_1\ne\lambda_2$. Then, for every $m\in\Z$, all gaps
with winding number $m$ are open in
$\Sigma_{\lambda_1,\lambda_2,\alpha}$, in the sense of
\eqref{eq:labels}.
\end{theorem}

The Diophantine case was proved in \cite{CL26}. We separate irrational
frequencies according to
\begin{align}
 \beta(\alpha)=\limsup_{n\to\infty}\frac{\log q_{n+1}}{q_n},
\end{align}
where $q_n$ are the continued-fraction denominators. The case $\beta=0$
also includes some non-Diophantine frequencies.

\medskip
Following the ideas of \cite{BorgniaFaust2026}, we can further extend the results of Theorem~\ref{thm:interior} to the critical case and the boundary case.

\begin{samepage}
\begin{theorem}[$L^*$]\label{thm:classification} For every
irrational $\alpha$ and every $x\in\T$, the following assertions hold for $0<\lambda_1,\lambda_2\leq 1$.
\begin{enumerate}[label=(\roman*)]
\item If $0<\lambda_1,\lambda_2<1$, all gaps with winding number
$m$ are open for every $m\in\Z$, whether or not $\lambda_1=\lambda_2$.
\item If $\max\{\lambda_1,\lambda_2\}=1$, all gaps with odd winding
number are open, and no gap with even winding number is open.
\end{enumerate}
\end{theorem}
\end{samepage}

Compared to the AMO case, the closure of all gaps with even winding number at boundary couplings is a new phenomenon. In the rest of the paper, we distinguish the argument into two separate parts: the proof for the interior noncritical case and the proof for the critical and boundary case.

Section~\ref{Sec:one} establishes gap opening in the noncritical interior, proving Theorem~\ref{thm:interior}. Section~\ref{Sec:two} treats boundary couplings: we establish the obstruction to opening gaps with even winding number and show that all gaps with each fixed odd winding number are open when one coupling equals $1$ and the other is sufficiently small. In Section~\ref{Sec:montonicity}, we prove that gap heights are nondecreasing as the smaller coupling increases with the larger coupling held fixed, allowing us to continue these gaps to the critical regime and the boundary corner. Finally, Section~\ref{sec:fullproof} combines the interior seeds of Theorem~\ref{thm:interior}, the boundary results, and height persistence to prove Theorem~\ref{thm:classification}.

\section{The noncritical case}~\label{Sec:one}
In this section, we discuss the case with couplings $0<\lambda_1\neq \lambda_2<1$. Our strategy follows the ideas of \cite{AYZ24} and takes their quantitative almost reducibility as substantial input.
For an analytic matrix function, $\|\cdot\|_h$ denotes the supremum of
its Euclidean operator norm on $|\operatorname{Im}x|<h$. All constants
may depend on the fixed couplings and on the analytic strip. They are
independent of the selected denominator scale unless stated otherwise.
Constants in the normal forms may also depend on the fixed frequency,
resonance, and spectral parameter.

\subsection{The real cocycle and its positive twist}\label{subsec:real-cocycle}
Let us first introduce a few notations for simplicity of our statement. Denote
\begin{align}
 c=\lambda_1',\qquad b(x)=\lambda_2\sin(2\pi x),\qquad
 p(x)=\sqrt{1-b(x)^2},\qquad
 J=\begin{pmatrix}0&-1\\1&0\end{pmatrix}.
\end{align}
Choose the analytic branch of $p$ positive on the real axis, on a
sufficiently small fixed strip. The two-step Szeg\H{o} matrix is
\begin{align}
 \mathcal S_z(x)=\frac1{\lambda_1p(x)}
 \begin{pmatrix}
 z+cb(x)&-b(x)-cz^{-1}\\
 -b(x)-cz&z^{-1}+cb(x)
 \end{pmatrix}.
\end{align}
With
\begin{align}
 U_C=\frac1{\sqrt2}\begin{pmatrix}1&i\\1&-i\end{pmatrix},
 \qquad z=e^{it},
\end{align}
its real form is
\begin{align}\label{eq:real-cocycle}
 A_t(x)=U_C^{-1}\mathcal S_{e^{it}}(x)U_C
 =\frac1{\lambda_1p(x)}
 \begin{pmatrix}
 (1-c)(\cos t-b(x))&-(1-c)\sin t\\
 (1+c)\sin t&(1+c)(\cos t+b(x))
 \end{pmatrix}.
\end{align}
It has determinant one, is real on the real axis, and is homotopic to
an $x$-independent matrix by replacing $b$ with $\vartheta b$ and
$p$ with $\sqrt{1-\vartheta^2b^2}$, $0\le\vartheta\le1$.
This homotopy stays in $\mathrm{SL}(2,\R)$. In particular, its
degree in $x$ is zero.
Direct differentiation gives
\begin{align}\label{eq:positive-twist}
 A_t^{-1}\partial_tA_t=J\mathcal Q_t,
 \qquad
 \mathcal Q_t(x)=\frac1{p(x)^2}
 \begin{pmatrix}
 1-b(x)\cos t&b(x)\sin t\\
 b(x)\sin t&1+b(x)\cos t
 \end{pmatrix}.
\end{align}
The eigenvalues of $\mathcal Q_t(x)$ are
$1/(1+|b(x)|)$ and $1/(1-|b(x)|)$. Hence, uniformly in real $x,t$,
\begin{align}\label{eq:twist-bounds}
 \frac1{1+\lambda_2}I\preceq\mathcal Q_t(x)
 \preceq\frac1{1-\lambda_2}I,
\end{align}
where we used $A\preceq B$ to mean the Loewner order: $\langle v,Av\rangle\leq \langle v,Bv\rangle$ for any $v\in\R^2$.
Thus the positive-twist hypothesis used below is verified for this
specific cocycle, with the spectral parameter measured in radians.

We use the standard spectral facts for this normalization: uniform
hyperbolicity is equivalent to $e^{it}\notin\Sigma$; the rotation number
is continuous, with a lift increasing by one when $t$ increases by
$2\pi$; and its gap labels agree with the two-step Szeg\H{o} convention.
The spectral characterization follows from \cite[Theorem 4.11]{CFO23},
using the nonsingular interior gauge, and the rotation-number and duality
conventions are discussed in \cite[Sections 3--4]{CL26}.
For $\lambda_1>\lambda_2$, $(\alpha,A_t)$ is subcritical at spectral
parameters by \cite[Theorem 2.9(a)]{CFO23}.

\subsection{A quantitative gap-opening lemma}
We use projective degree: $\deg B=m$ means that the line spanned
by the first column of $B$ winds $m$ times in $\mathbb{RP}^1$.
A determinant-one lift then satisfies $B(x+1)=(-1)^mB(x)$.
With this convention, conjugation by $B$ changes the rotation
number by $m\alpha/2$ modulo one. Multiplication of the cocycle
by $-I$ changes it by $1/2$. These are distinct operations;
the sign $\sigma$ below records only the latter.
\begin{lemma}[$L^*$]\label{lem:opening}
Let $A_t:\T\to\mathrm{SL}(2,\R)$ be a continuous cocycle family,
$C^2$ in $t$, satisfying
$A_t^{-1}\partial_tA_t=JQ_t$ with symmetric matrices
$q_0I\preceq Q_t\preceq q_1I$, where $q_0>0$.
Suppose that a continuous projective conjugacy $B$ has a periodic or
antiperiodic determinant-one lift and that, for some
$\sigma\in\{1,-1\}$,
\begin{align}
 \sigma B(x+\alpha)^{-1}A_{t_0}(x)B(x)=U_d+E(x),
 \qquad U_d=\begin{pmatrix}1&d\\0&1\end{pmatrix},
 \qquad 0<|d|\le1.
\end{align}
Assume $\|B\|_0\le K$, $K\ge2$. There exist constants $c_0,c_1>0$,
depending only on $q_0,q_1$, such that
\begin{align}\label{eq:openingcondition}
 \|E\|_0\le c_0|d|K^{-4}
\end{align}
implies uniform hyperbolicity for every $t=t_0+s$ with
\begin{align}\label{eq:opening-window}
 \sgn s=\sgn d,\qquad
 \tfrac12c_1|d|K^{-2}<|s|<c_1|d|K^{-2}.
\end{align}
If $\deg B=m$, the rotation number on this interval is
\begin{align}
 \rho(t)=\frac{m\alpha}{2}+\frac{1-\sigma}{4}\pmod1.
\end{align}
In particular, if $\rho(t_0)$ equals this value, the interval has
the same rotation number as $t_0$. The parameter window is assumed
to lie in the interval where the family is defined.
\end{lemma}

\begin{proof}
All matrices below are evaluated at $x$ unless indicated otherwise.
Set
\begin{align}
 \widetilde A_s=\sigma B(x+\alpha)^{-1}A_{t_0+s}(x)B(x),
 \qquad R_s=B^{-1}A_{t_0}^{-1}A_{t_0+s}B.
\end{align}
Then
\begin{align}\label{eq.perturbConj1}
 \widetilde A_s=(U_d+E)R_s,\qquad
 R_0=I,\qquad \partial_sR_s=R_sJS_s,
 \qquad S_s=B^TQ_{t_0+s}B.
\end{align}
Here $B^{-1}J=JB^T$. Since $\det B=1$,
$\|B^{-1}\|=\|B\|$, and the singular-value bounds give
\begin{align}\label{eq.definiteCompar}
 q_0K^{-2}I\preceq S_s\preceq q_1K^2I.
\end{align}
No ordering of the nonsymmetric matrix $B$ itself is used.
For $|s|\le c_1|d|K^{-2}$, Gronwall's inequality yields
\begin{align}
 \|R_s\|\le e^{q_1K^2|s|},\qquad
 \|R_s-I\|\le e^{q_1K^2|s|}-1\le Cc_1|d|.
\end{align}
Choose $c_1$ sufficiently small, and write
\begin{align}
 S_s=\begin{pmatrix}a_s&b_s\\b_s&c_s\end{pmatrix},
 \qquad R_s=\begin{pmatrix}p_s&q_s\\r_s&u_s\end{pmatrix}.
\end{align}
In particular, $u_s\ge3/4$. The bottom row of \eqref{eq.perturbConj1}
satisfies
\begin{align}
 r_s'=a_su_s-b_sr_s,\qquad u_s'=b_su_s-c_sr_s.
\end{align}
For $z_s=r_s/u_s$ this implies
\begin{align}
 z_s'=a_s-2b_sz_s+c_sz_s^2
 = (1,-z_s)S_s(1,-z_s)^T\ge q_0K^{-2},\qquad z_0=0.
\end{align}
Integration in either direction from zero gives
\begin{align}\label{eq.estRs}
 \sgn r_s=\sgn s,\qquad
 |r_s|\ge\tfrac34q_0K^{-2}|s|.
\end{align}
On the window \eqref{eq:opening-window}, therefore,
\begin{align}
 \sgn r_s=\sgn d,\qquad
 |r_s|\ge\tfrac38c_1q_0|d|K^{-4}.
\end{align}
On the other hand,
\begin{align}
 U_dR_s=\begin{pmatrix}p_s+dr_s&q_s+du_s\\r_s&u_s\end{pmatrix},
 \qquad |q_s+du_s-d|\le Cc_1|d|.
\end{align}
After reducing $c_1$, its upper-right entry has sign $d$ and modulus
at least $3|d|/4$, and both diagonal entries exceed $1/2$.
Moreover,
\begin{align}
 \|ER_s\|\le Cc_0|d|K^{-4}.
\end{align}
Choose $c_0$ after $c_1$ so that this perturbation is less than half
of each required positive margin. Then $\widetilde A_s$ has positive
diagonal entries and off-diagonal entries of sign $d$, with
\begin{align}
 |(\widetilde A_s)_{21}|\ge\tfrac3{16}c_1q_0|d|K^{-4},
 \qquad |(\widetilde A_s)_{12}|\ge\tfrac12|d|.
\end{align}
Conjugating by $\operatorname{diag}(1,\sgn d)$ makes every entry
strictly positive. For fixed $d,K$, these matrices form a compact
subset of the strictly positive determinant-one matrices. They map
the closed positive projective cone uniformly into its interior;
the strict cone criterion gives a dominated splitting. Determinant
one upgrades this splitting to uniform hyperbolicity.

The invariant expanding direction lies in a fixed quadrant, so it
has degree zero, and the associated vector is mapped to a positive
multiple of the next vector. Thus the near-$U_d$ lift of the
rotation number of $\widetilde A_s$ is zero modulo one. Restoring
the sign $\sigma$ adds $0$ when $\sigma=1$ and $1/2$ when
$\sigma=-1$, that is, $(1-\sigma)/4$ modulo one. Conjugation by
a degree-$m$ projective map adds $m\alpha/2$. This proves the
stated rotation-number formula as well as uniform hyperbolicity.
\end{proof}

\subsection{Quantitative almost reducibility}
The one-step transfer matrix of the eigenvalue equation acting on $(u_n^+,u_{n-1}^-)^T$ is
\begin{align}\label{eq:physicaltransfer}
 T_z(x)=\frac1{\varrho(x)}
 \begin{pmatrix}
 \lambda_1^{-1}(z^{-1}+2cb(x)+c^2z)&-b(x)-cz\\
 -b(x)-cz&\lambda_1z
 \end{pmatrix},
\end{align}
where $\varrho(x)=\lambda_2\cos2\pi x-i\lambda_2'.$
Here $c,b,p,U_C$ have the meanings fixed above. Let 
\begin{align}\label{eq:spin-flip}
    R_0=\begin{pmatrix}1&0\\-c&\lambda_1\end{pmatrix},\quad
X_0=\begin{pmatrix}0&1\\1&0\end{pmatrix},\quad \text{and} \quad
L=(X_0R_0)^{-1}U_C
\end{align}
Direct multiplication gives
\begin{align}\label{eq:physicalgauge}
 T_z(x)=\eta(x)LA_t(x)L^{-1},\qquad
 \eta(x)=\frac{p(x)}{\varrho(x)}.
\end{align}
On the real torus, $|\eta|=1$. With
$r=\lambda_2/(1+\lambda_2')<1$,
\begin{align}\label{eq:phasefactor}
 \varrho(x)=\frac{-i}{1+r^2}
 (1+i r e^{2\pi i x})(1+i r e^{-2\pi i x}).
\end{align}
Consequently $\eta=i e^{i\varphi}$ for a real analytic, periodic,
mean-zero function $\varphi$. This scalar phase must be retained when
Fourier-transforming a Bloch frame.

\begin{lemma}[$L^*$]\label{lem:gauge}
Let $\varphi$ be real analytic and mean zero. For consecutive convergent
denominators $q,Q$ with $Q>2q$, define
\begin{align}
 u_q(x)=\sum_{0<|k|<Q/2}
 \frac{\widehat\varphi(k)}{e^{2\pi i k\alpha}-1}e^{2\pi i kx}.
\end{align}
On a fixed smaller strip of width $s>0$, there are constants $c,C>0$ such
that
\begin{align}\label{eq:gaugebounds}
 \|u_q\|_s\le C(q+Qe^{-cq}),\qquad
 \Vert\varphi-u_q(\cdot+\alpha)+u_q\Vert_s\le Ce^{-cQ}.
\end{align}
If $\beta(\alpha)=0$, the full cohomological equation
$u(x+\alpha)-u(x)=\varphi(x)$ has a real analytic solution on a smaller
positive strip.
\end{lemma}

\begin{proof}
Best approximation bounds the inverse Fourier divisors by $Cq$ for
$0<|k|<q$, and by $CQ$ for $0<|k|<Q$. Split the exponentially decaying
Fourier series at $q$ to obtain the first estimate. The omitted tail gives
the second, after reducing $c$. Conjugate symmetry makes $u_q$ real.
When $\beta=0$, inverse divisors grow subexponentially, so exponential
Fourier decay gives convergence of the full series on a smaller strip.
\end{proof}
In particular, $g_q=e^{i u_q}$ is unimodular on the real torus and
\begin{align}
 \eta\,\frac{g_q}{g_q(\cdot+\alpha)}
 =i+O_s(e^{-cQ}),\qquad
 \Vert g_q^{\pm1}\Vert_s\le e^{C(q+Qe^{-cq})}.
\end{align}

\smallskip\noindent\emph{External resonant normal-form input.}
We use the following quantitative and exact resonant normal-form inputs
from \cite[Theorems 3.2--3.3]{AYZ24}, in the finite-rate form specified below.
They apply to the normalized analytic subcritical real cocycle, with its
actual rotation number and full projective degree. Qualitative almost reducibility alone is not the quantitative input below.
The literature input at this point is precisely the cited AYZ normal-form result.

Fix an analytic subcritical cocycle $A$ with rotation number
$\rho_*$ satisfying $2\rho_*\equiv m\alpha\pmod1$, and set
\begin{align}
 \sigma=\exp\bigl(2\pi i\rho_*-\pi im\alpha\bigr)\in\{1,-1\}.
\end{align}
This sign is determined by $A$ and $m$; it is not an additional
gap label. A projective analytic conjugation of degree $m$ reduces
$A$ to $\sigma U_d$ when $\beta(\alpha)=0$; compare \cite[Theorem 3.3]{AYZ24}.
When $\beta(\alpha)>0$, by \cite[Theorem 3.2]{AYZ24} one can choose selected
consecutive denominators $q_n,Q_n=q_{n+1}$ and conjugations $B_n$ of degree $m$
so that
\begin{align}\label{eq:nf}
 \sigma B_n(x+\alpha)^{-1}A(x)B_n(x)=U_{d_n}+E_n(x),
\end{align}
\begin{align}\label{eq:nfest}
 \Vert B_n\Vert_s\le e^{2Q_n\tau_n},\,
 |d_n|\le e^{-Q_n\tau_n},\,
 \Vert E_n\Vert_s\le e^{-Q_n\Delta_n}.
\end{align}
The strip width $s>0$ is independent of the selected scale.
Here $n$ always denotes the original continued-fraction index restricted
to an increasing selected subsequence; it is not renumbered after selection.
When using a source convention with frequency in $[0,1)$, first reduce
$\alpha$ modulo one and retain the corresponding projective sign.
Choose
$0<b_*<\beta(\alpha)$ when $\beta<\infty$, and any finite $b_*>0$
when $\beta=\infty$. The selected consecutive denominators satisfy
$Q_n\ge e^{b_*q_n}$. In the notation of \cite[Section 3]{AYZ24},
$\tau_n=\varepsilon_n^{1/4}\asymp e^{-\delta' q_n/8}$ and
$\Delta_n=h_*/2^n$; its strips have a positive limiting width.
Choose the initial parameter $\delta'>0$ sufficiently small so that
$\kappa=\delta'/8<b_*$ and $\kappa<c_{\mathrm g}$, where
$c_{\mathrm g}$ is a scalar-gauge Fourier decay rate. This last choice
is compatible with shrinking $s$: the scalar phase is analytic on a
fixed initial strip, so its Fourier tail has a decay rate bounded
below on any strip of at most half that width. Then
\begin{align}
 \tau_n\asymp e^{-\kappa q_n},\qquad
 Q_n\tau_n\longrightarrow\infty,\qquad
 \Delta_n/\tau_n\longrightarrow\infty.
\end{align}
The case $\beta=\infty$ uses the same finite lower rate $b_*$.
Indeed, the selected rational approximants satisfy
$|\alpha-p_n/q_n|<1/(q_nQ_n)\le e^{-b_*q_n}/q_n$.
For $\beta=\infty$, the use of \cite[Section~3]{AYZ24} is in
this finite-rate form: only the chosen finite lower rate $b_*$ is used
in the approximation estimate, while $q_n,Q_n$ retain their actual values.
The existence of the normal forms with the common strip and bounds
\eqref{eq:nf}--\eqref{eq:nfest}, including this infinite-$\beta$ use,
is an external analytic input. The subsequent gauge absorption and
selected-scale limits are checked consequences of those data; they do
not construct the data.

The sign $\sigma$ is fixed by $\rho_*$ and $m$ at every selected
scale. For $\beta=0$, it is retained in the exact reduction of
\cite[Theorem 3.3]{AYZ24}. To identify the degree in that case,
write $k=\deg B$. The conjugacy-degree formula and the parabolic
normal form give $2\rho_*\equiv k\alpha\pmod1$.
Comparing with $2\rho_*\equiv m\alpha\pmod1$ and using
irrationality of $\alpha$ gives $k=m$.

A lift of a degree-\(m\) conjugation satisfies
\(
B(x+1)=(-1)^mB(x).
\)
Let \(u_q\) be given by Lemma \ref{lem:gauge} and set
\(
g_q=e^{iu_q}.
\)
We then introduce the periodic physical frame
\begin{align}\label{eq:frame}
Z_q(x)=e^{-\pi i mx}g_q(x)LB_q(x).
\end{align}
The factor \(e^{-\pi i mx}\) compensates for the possible antiperiodicity of the lift \(B_q\), while \(g_q\) removes the \(x\)-dependent scalar phase in \eqref{eq:physicalgauge}. More precisely, writing the scalar-gauge remainder separately, we obtain
\begin{align}\label{eq:transferEq}
Z_q(x+\alpha)^{-1}T_z(x)Z_q(x)
=
\varkappa\bigl(U_{d_q}+\widehat E_q(x)\bigr),
\qquad
\|\widehat E_q\|_s\le C\bigl(\|E_q\|_s+e^{-c_{\mathrm g}Q}\bigr),
\end{align}
where
\begin{align}\label{eq:blochphase}
\varkappa=i\sigma e^{\pi im\alpha}
=i e^{2\pi i\rho_*}=e^{2\pi i\xi},
\qquad \xi=\tfrac14+\rho_*\pmod1.
\end{align}
Thus the whole frame carries the common scalar multiplier \(\varkappa\).

Since $\kappa<c_{\mathrm g}$ and $\kappa<b_*$,
\begin{align}
q+Qe^{-c_{\mathrm g}q}=o(Q\tau_q),
\end{align}
the bounds for \(B_q\) and \(u_q\) imply, after enlarging a fixed constant,
\begin{align}
\|Z_q\|_s\le \mathcal M_q,
\qquad
\mathcal M_q=e^{C_0Q\tau_q}.
\end{align}
Moreover,
\begin{align}\label{eq:gramframe}
\int_{\mathbb T}|Z_q(x)v|^2\,\dd x
\ge
\mathcal M_q^{-2}|v|^2,
\qquad v\in\mathbb C^2.
\end{align}
Indeed, the scalar factors have modulus one on the real axis, and
$|LB_q(x)v|\ge \|L^{-1}\|^{-1}e^{-2Q\tau_q}|v|$ there.
Increasing $C_0$ absorbs this fixed factor and proves
\eqref{eq:gramframe}.

\subsection{Fourier duality}\label{sec:duality}
For a column profile $v$ of $Z_q$, define
\begin{align}
 u_n^+=\varkappa^nv_1(\theta+n\alpha),\qquad
 u_n^-=\varkappa^{n+1}v_2(\theta+(n+1)\alpha).
\end{align}
Then $X_n=(u_n^+,u_{n-1}^-)^T=\varkappa^nv(\theta+n\alpha)$.
The transfer equation \eqref{eq:transferEq} is exact only if 
$T_zv=\varkappa v(\cdot+\alpha)$. For the two-column frame in general,
the defect includes both the shear of $U_{d_q}$ and error $\widehat E_q$.

Writing $u_n=\varkappa^n f(\theta+n\alpha)$, define the shift operator $\mathscr S_{\lambda_1,\varkappa}$ as
\begin{align}
(\mathscr S_{\lambda_1,\varkappa}f)(x)
=
\begin{pmatrix}
\lambda_1\varkappa^{-1}f_+(x-\alpha)-\lambda_1'f_-(x)\\
\lambda_1\varkappa f_-(x+\alpha)+\lambda_1'f_+(x)
\end{pmatrix},
\end{align}
and let the coin operator $Q_{\lambda_2}$ act by multiplication by
$Q_{\lambda_2}(x)$. Then define the walk operator $\mathscr W_\varkappa=\mathscr S_{\lambda_1,\varkappa}Q_{\lambda_2}$. With these definitions,
\begin{align}
(W_{\lambda_1,\lambda_2,\alpha,\theta}u)_n
=
\varkappa^n
(\mathscr W_\varkappa f)(\theta+n\alpha).
\end{align}
Appendix~\ref{app:duality} verifies that the unitary operator $\mathcal U=S_{\lambda_2}K_F\mathcal F$ from
$L^2(\T;\C^2)$ to $\ell^2(\Z;\C^2)$ satisfies 
\begin{align}\label{eq:walkOperConj}
    \mathcal{U}\mathscr{W}_{\varkappa}=W_{\lambda_2,\lambda_1,-\alpha,-\xi}\mathcal{U}.
\end{align}

\subsection{Walk and transfer residuals}
We now relate errors in the transfer equation to errors in the walk
eigenvalue equation. Write
\begin{align}
Q_n=(q_n^{ij})_{i,j=1}^2,
\qquad
X_n=
\begin{pmatrix}
u_n^+\\
u_{n-1}^-
\end{pmatrix},
\qquad
R_n=X_{n+1}-T_nX_n.
\end{align}
Thus $R_n$ measures the failure of $u$ to satisfy the transfer
recurrence. Substituting this recurrence into the two components
of $(Q-zS_{\lambda_1}^*)u$ gives
\begin{align}\label{eq:localresidual}
\bigl((Q-zS_{\lambda_1}^*)u\bigr)_n
=C_nR_n,
\qquad
C_n=
\begin{pmatrix}
-z\lambda_1&q_n^{12}-z\lambda_1'\\
0&q_n^{22}
\end{pmatrix}.
\end{align}
In particular, the expression on the left at site $n$ depends
only on the transfer residual $R_n$ at that site.
For fixed interior couplings $0<\lambda_1,\lambda_2<1$,
the matrices $C_n$ and $C_n^{-1}$ are uniformly bounded:
indeed, $|z|=1$ and
\begin{align}
|q_n^{22}|
=
\left|
\lambda_2\cos\bigl(2\pi(\theta+n\alpha)\bigr)
-i\lambda_2'
\right|
\ge \lambda_2'>0.
\end{align}
Since $W=S_{\lambda_1}Q$, the walk residual is
\begin{align}
(W-z)u=S_{\lambda_1}(CR),
\qquad
(CR)_n=C_nR_n.
\end{align}
Thus passing from a transfer residual to a walk residual requires
only multiplication by $C_n$, followed by the nearest-neighbor
shift $S_{\lambda_1}$. Both operations preserve exponential
decay, up to a constant.

We apply this observation to an analytic conjugation $Z$
satisfying
\begin{align}
T_z(x)Z(x)
=\varkappa Z(x+\alpha)\bigl(I+\mathcal E(x)\bigr).
\end{align}
Here $\mathcal E$ is the error relative to an exact scalar
reduction; the corresponding transfer error is
\begin{align}
T_z(x)Z(x)-\varkappa Z(x+\alpha)
=\varkappa Z(x+\alpha)\mathcal E(x).
\end{align}
Let $\Phi$ be the associated same-site frame,
\begin{align}
\Phi(x)=
\begin{pmatrix}
Z_{11}(x)&Z_{12}(x)\\
\varkappa Z_{21}(x+\alpha)&
\varkappa Z_{22}(x+\alpha)
\end{pmatrix},
\end{align}
and define its walk residual, column by column, by
\begin{align}\label{eq:operIdty}
\mathscr R=(\mathscr W_\varkappa-z)\Phi.
\end{align}
If $\|Z\|_s\le\mathcal M$, the preceding residual conversion
gives, on a fixed smaller strip $0<s'<s$,
\begin{align}
\|\mathscr R\|_{s'}
\le C\mathcal M\|\mathcal E\|_s.
\end{align}

Now apply Fourier duality to each column and denote
\begin{align}
\Psi=\mathcal U\Phi,
\qquad
W_{\mathrm{dual}}
=W_{\lambda_2,\lambda_1,-\alpha,-\xi}.
\end{align}
The operator identity \eqref{eq:operIdty} implies $(W_{\mathrm{dual}}-z)\Psi=\mathcal U\mathscr R.$
Analyticity on a fixed strip gives constants $a,C>0$ such that
\begin{align}
\|\Psi_n\|
\le C\mathcal M e^{-a|n|},
\qquad
\bigl\|((W_{\mathrm{dual}}-z)\Psi)_n\bigr\|
\le C\mathcal M\|\mathcal E\|_s e^{-a|n|}.
\end{align}
Moreover, the same-site conversion and $\mathcal U$ are unitary,
so the  lower bound \eqref{eq:gramframe} is preserved:
\begin{align}
\sum_{n\in\mathbb Z}|\Psi_n b|^2
=
\int_{\mathbb T}|Z(x)b|^2\,dx
\ge \mathcal M^{-2}|b|^2,
\qquad b\in\mathbb C^2.
\end{align}
We therefore obtain two quantitatively independent dual vectors
with exponential decay and exponentially decaying eigenvalue
residuals controlled by $\mathcal M\|\mathcal E\|_s$.
For the normal form
$T_zZ_q=\varkappa(Z_q)_+(U_{d_q}+\widehat E_q)$, the error used here is
\begin{align}
\mathcal E=U_{d_q}-I+\widehat E_q,
\end{align}
so its size includes both the shear $d_q$ and the remainder $E_q$.
The reverse passage, needed on the dual side, is equally important:
\begin{align}\label{eq:reverse-residual}
 R_n=C_n^{-1}\bigl(S_{\lambda_1}^*(W-z)u\bigr)_n.
\end{align}
Thus exponentially decaying walk residuals give transfer residuals
with the same decay exponent, up to a fixed constant. The staggered
state $X_n=(u_n^+,u_{n-1}^-)^T$ preserves the full $\ell^2$ Gram
matrix by reindexing its second row.

\subsection{A quantitative Wronskian obstruction}
We will need the following lemma to show that in the reducible case, $d\neq 0$ and in the almost reducible case, $d_n$ is quantitatively bounded from below.
\begin{lemma}[$L^*$]\label{lem:wronskian}
Fix $a>0$ and $M_0\ge1$. Suppose
\begin{align}
 X_{n+1}=T_nX_n+R_n,\qquad
 |\det T_n|=1,\quad\Vert T_n^{\pm1}\Vert\le M_0,
\end{align}
where the columns of $X_n$ obey
\begin{align}
 \Vert X_n\Vert\le\mathcal M e^{-a|n|},\quad
 \Vert R_n\Vert\le\varepsilon\mathcal M e^{-a|n|},\quad
 \sum_n\Vert X_nv\Vert^2\ge\mathcal M^{-2}|v|^2,
 \quad \mathcal M\ge2.
\end{align}
Then $\varepsilon\ge c\mathcal M^{-P}$, with $c,P>0$ depending only on
$a,M_0$.
\end{lemma}

\begin{proof}
We may assume $0\le\varepsilon\le1$, since the conclusion for
$\varepsilon>1$ follows by choosing $c\le1$.

Write $X_n=[x_n^{(1)},x_n^{(2)}]$ and
$R_n=[r_n^{(1)},r_n^{(2)}]$, and set $w_n=\det X_n$.
Expanding the determinant in
$X_{n+1}=T_nX_n+R_n$ gives
\begin{align}
\begin{aligned}
w_{n+1}-(\det T_n)w_n
={}&\det(T_nx_n^{(1)},r_n^{(2)})
+\det(r_n^{(1)},T_nx_n^{(2)}) +\det(r_n^{(1)},r_n^{(2)}).
\end{aligned}
\end{align}
Using $|\det(u,v)|\le |u|\,|v|$ and $\varepsilon\le1$, we obtain
\begin{align}
\left|w_{n+1}-(\det T_n)w_n\right|
\le C\varepsilon\mathcal M^2e^{-2a|n|}.
\end{align}
Since $|\det T_n|=1$, iteration from $n$ to $N>n$ yields
\begin{align}
|w_n|
\le |w_N|
+C\varepsilon\mathcal M^2
\sum_{k=n}^{N-1}e^{-2a|k|}.
\end{align}
The decay of $X_N$ implies $w_N\to0$ as $N\to+\infty$.
Consequently,
\begin{align}\label{eq:wronskian-small}
|\det X_n|\le C\varepsilon\mathcal M^2
\qquad\text{for every }n\in\mathbb Z.
\end{align}
Let $L=\left\lceil A_a\log\mathcal M\right\rceil,$
where $A_a$ is sufficiently large that
\begin{align}
\mathcal M^2\sum_{|n|>L}e^{-2a|n|}
=
\frac{2\mathcal M^2e^{-2a(L+1)}}{1-e^{-2a}}
\le\frac{1}{4\mathcal M^2}.
\end{align}
Such a choice is possible uniformly for $\mathcal M\ge2$.
For every unit vector $v$, the decay assumption therefore gives
\begin{align}
\sum_{|n|>L}|X_nv|^2\le\frac{1}{4\mathcal M^2}.
\end{align}
Combining this with the Gram lower bound, we find
\begin{align}\label{eq:window-gram}
\sum_{|n|\le L}|X_nv|^2
\ge\frac{3}{4\mathcal M^2},
\qquad |v|=1.
\end{align}

Let $N_L=2L+1$. Applying \eqref{eq:window-gram} to a fixed
unit vector, for example $e_1$, shows that some
$n_0\in[-L,L]$ satisfies
\begin{align}
s_{\max}(X_{n_0})
=\|X_{n_0}\|
\ge |X_{n_0}e_1|
\ge\frac{\sqrt{3}}{2\mathcal M\sqrt{N_L}}.
\end{align}
Hence
\eqref{eq:wronskian-small} implies
\begin{align}
s_{\min}(X_{n_0})
=\frac{|\det X_{n_0}|}{s_{\max}(X_{n_0})}
\le C\varepsilon\mathcal M^3\sqrt{N_L}.
\end{align}
Choose a unit singular vector $v$ corresponding to
$s_{\min}(X_{n_0})$, and keep this vector fixed at every site.
Define
\begin{align}
y_n=X_nv,\qquad f_n=R_nv.
\end{align}
Then
\begin{align}
y_{n+1}=T_ny_n+f_n,
\qquad
|f_n|\le\varepsilon\mathcal M,
\qquad
|y_{n_0}|\le C\varepsilon\mathcal M^3\sqrt{N_L}.
\end{align}

We now propagate this small initial value across $[-L,L]$.
Forward iteration uses $\|T_n\|\le M_0$, while backward
iteration uses
\begin{align}
y_n=T_n^{-1}y_{n+1}-T_n^{-1}f_n
\end{align}
and $\|T_n^{-1}\|\le M_0$.
In either direction, if $d=|n-n_0|\le2L$, these bounds give
\begin{align}
|y_n|
\le M_0^d\bigl(|y_{n_0}|+d\varepsilon\mathcal M\bigr).
\end{align}
Therefore, for every $|n|\le L$,
\begin{align}
\begin{aligned}
|y_n|
&\le
M_0^{2L}
\left(
C\varepsilon\mathcal M^3\sqrt{N_L}
+2L\varepsilon\mathcal M
\right)\le C\varepsilon\mathcal M^3N_LM_0^{2L}.
\end{aligned}
\end{align}
Taking the $\ell^2$ norm over the window yields
\begin{align}
\left(\sum_{|n|\le L}|X_nv|^2\right)^{1/2}
\le C\varepsilon\mathcal M^3N_L^{3/2}M_0^{2L}.
\end{align}
On the other hand, \eqref{eq:window-gram} applies to this
same unit vector $v$, so
\begin{align}
\frac{\sqrt{3}}{2\mathcal M}
\le C\varepsilon\mathcal M^3N_L^{3/2}M_0^{2L}.
\end{align}
It follows that $\varepsilon \ge c\mathcal M^{-4}N_L^{-3/2}M_0^{-2L}.$
Finally, $L\le C_a\log\mathcal M$ and
$N_L\le C_a\log\mathcal M$ for $\mathcal M\ge2$.
Thus there exists $P>0$, depending only on $a$ and $M_0$,
such that $N_L^{3/2}M_0^{2L}\le C\mathcal M^{P-4}.$
Substitution gives the claimed estimate \begin{align}\varepsilon\ge c\mathcal M^{-P}.
\end{align}
\end{proof}

\subsection{Proof of Theorem~\ref{thm:interior}}
\begin{proof}
Aubry duality preserves the spectrum and the density-of-states gap
labels; see \cite[Corollary 2.6]{CFO23} and \cite[Section 4]{CL26}.
It therefore suffices to treat $\lambda_1>\lambda_2$.
Fix $m\in\Z$ and an arbitrary value $\rho_*\in\R/\Z$ satisfying
$2\rho_*\equiv m\alpha\pmod1$. If $\rho_*$ already occurs on an
open gap, there is nothing to prove. Otherwise, continuity and the
spectral interpretation of the rotation number give a spectral
parameter $t_0$ with $\rho(t_0)=\rho_*$ modulo one. By the spectral
facts stated after \eqref{eq:twist-bounds}, $(\alpha,A_{t_0})$ is
subcritical. Use the normal forms above with the sign
\begin{align}
 \sigma=e^{2\pi i\rho_*-\pi im\alpha},\qquad
 \rho_*=\frac{m\alpha}{2}+\frac{1-\sigma}{4}\pmod1.
\end{align}

\smallskip
\noindent\textit{The common dual estimates.}
Let $Z$ be one of the periodic analytic physical frames constructed
above. Apply the duality transform $\mathcal U$ from Section~\ref{sec:duality}
to each column of its associated same-site frame, and denote the
resulting dual sequences by $\psi^{(1)}$ and $\psi^{(2)}$. Let
\begin{align}
 X_n^{\mathrm d}
 =
 \begin{pmatrix}
 \psi_n^{(1),+} & \psi_n^{(2),+}\\
 \psi_{n-1}^{(1),-} & \psi_{n-1}^{(2),-}
 \end{pmatrix}.
\end{align}
Let $T_n^{\mathrm d}$ be the matrix in
\eqref{eq:physicaltransfer}, with $\lambda_1$ and $\lambda_2$
interchanged and $x=-\xi-n\alpha$, at spectral parameter
$z=e^{it_0}$. Let
 $\varrho_n^{\mathrm d}
 =\lambda_1\cos\bigl(2\pi(-\xi-n\alpha)\bigr)-i\lambda_1'.$
The explicit formula for $T_n^{\mathrm d}$ gives
 $\det T_n^{\mathrm d}=\frac{\overline{\varrho_n^{\mathrm d}}}
        {\varrho_n^{\mathrm d}}.$
Since $|\varrho_n^{\mathrm d}|\ge\lambda_1'>0$, the entries
of $T_n^{\mathrm d}$ are uniformly bounded in $n$.
Moreover, $|\det T_n^{\mathrm d}|=1$, so its inverse is
uniformly bounded as well. Thus there is a constant $M_0$
such that
\begin{align}
 |\det T_n^{\mathrm d}|=1,
 \qquad
 \bigl\|(T_n^{\mathrm d})^{\pm1}\bigr\|\le M_0
 \quad\text{for all }n\in\mathbb Z.
\end{align}
The residual conversion \eqref{eq:reverse-residual}, Fourier decay,
and the unitary same-site map give, after absorbing fixed constants
in $\mathcal M$,
\begin{align}\label{eq:dual-estimates}
 \|X_n^{\mathrm d}\|\le\mathcal M e^{-a|n|},\quad
 \|R_n^{\mathrm d}\|\le C\|\mathcal E\|_s\mathcal M e^{-a|n|},\quad
 \sum_{n\in\Z}|X_n^{\mathrm d}v|^2
 =\int_{\T}|Z(x)v|^2\,dx
 \ge\mathcal M^{-2}|v|^2.
\end{align}
Here $R_n^{\mathrm d}=X_{n+1}^{\mathrm d}-T_n^{\mathrm d}X_n^{\mathrm d}$,
and $a>0$ is fixed by the strip. We next distinguish the cases $\beta(\alpha)=0$ and $\beta(\alpha)>0$:

\smallskip
\noindent\textit{Case $\beta(\alpha)=0$.}
The exact reducibility input \cite[Theorem 3.3]{AYZ24}, followed by
the exact scalar gauge of Lemma~\ref{lem:gauge}, gives a periodic
analytic frame $Z$ satisfying
\begin{align}
 T_z(x)Z(x)=\varkappa Z(x+\alpha)U_d,\qquad z=e^{it_0}.
\end{align}
If $d=0$, then $\mathcal E=0$, so \eqref{eq:dual-estimates} satisfies
Lemma~\ref{lem:wronskian} with zero residual. This would imply
$0\ge c\mathcal M^{-P}>0$, a contradiction. Hence $d\ne0$.
If necessary, replace $B$ by
$B\operatorname{diag}(r,r^{-1})$ with $r\ge1$ so that the new shear
$d/r^2$ has modulus at most one. The degree $m$ and the sign $\sigma$ do not change.
The twist bounds \eqref{eq:twist-bounds} and
Lemma~\ref{lem:opening}, now with $E=0$, produce an open uniformly
hyperbolic interval on which $\rho(t)=\rho_*$ modulo one.
By the spectral characterization, it is contained in an open
spectral gap with the prescribed rotation-number value.

\smallskip
\noindent\textit{Case $\beta(\alpha)>0$.}
Use the normal forms \eqref{eq:nf}--\eqref{eq:nfest} and the periodic
physical frames \eqref{eq:frame}. Set
\begin{align}
 \mathcal M_q=e^{C_0Q\tau_q},\qquad
 \mathcal E_q=U_{d_q}-I+\widehat E_q.
\end{align}
The scalar gauge and the original real normal-form remainder are
kept separate, so
\begin{align}
 \|\mathcal E_q\|_s
 \le C\bigl(|d_q|+e^{-Q\Delta_q}+e^{-c_{\mathrm g}Q}\bigr).
\end{align}
Applying Lemma~\ref{lem:wronskian} to \eqref{eq:dual-estimates} gives
\begin{align}\label{eq:shear-obstruction}
 |d_q|+e^{-Q\Delta_q}+e^{-c_{\mathrm g}Q}
 \ge c\mathcal M_q^{-P}=c e^{-C_0P Q\tau_q}.
\end{align}
The scale relations are
\begin{align}
 \tau_q\longrightarrow0,\qquad Q\tau_q\longrightarrow\infty,
 \qquad \Delta_q/\tau_q\longrightarrow\infty.
\end{align}
For $A=C_0P$, they imply
\begin{align}
 e^{-Q\Delta_q+AQ\tau_q}
 &=\exp\!\left[-Q\tau_q(\Delta_q/\tau_q-A)\right]\longrightarrow0,\\
 e^{-c_{\mathrm g}Q+AQ\tau_q}
 &=\exp\!\left[-Q(c_{\mathrm g}-A\tau_q)\right]\longrightarrow0.
\end{align}
Absorbing both remainders in \eqref{eq:shear-obstruction} yields
\begin{align}\label{eq:shear-lower}
 |d_q|\ge c_*e^{-A Q\tau_q}>0
\end{align}
for all sufficiently large selected scales.

Finally take $K_q=\max\{2,e^{2Q\tau_q}\}$. The original real
remainder, rather than the complex physical-frame remainder, obeys
\begin{align}
 \frac{\|E_q\|_0K_q^4}{|d_q|}
 \le C\exp\!\left[-Q\Delta_q+(8+A)Q\tau_q\right]
 \longrightarrow0.
\end{align}
Also $|d_q|\le e^{-Q\tau_q}\le1$. Thus
Lemma~\ref{lem:opening} applies to the real family $A_t$, with the
uniform twist bounds already verified in \eqref{eq:twist-bounds}.
It produces a nonempty uniformly hyperbolic interval whose rotation
number is \emph{exactly} $\rho_*$ modulo one. This gives an open
spectral gap with the prescribed rotation-number value.
The argument applies to either solution of
$2\rho_*\equiv m\alpha\pmod1$ and to every $m\in\Z$.
Thus all the asserted gaps are open.
\end{proof}

\section{Boundary Couplings}~\label{Sec:two}
In this section, we discuss the gap opening when at least one of
$\lambda_1,\lambda_2$ equals $1$. We call this the
\emph{boundary coupling case}.

We will use the spectral criterion on the nonsingular edge
$(\lambda_1,\lambda_2)=(1,\lambda)$, $0<\lambda<1$.
Here $c=0$, $L=X_0U_C$ is invertible (see  \eqref{eq:physicalgauge} and the paragraph above it), and
$p(x)=|\varrho(x)|\ge\sqrt{1-\lambda^2}>0$ for real $x$.
The regularized transfer matrix of
\cite[equation~(4.19)]{CFO23} is
\begin{align}
 \mathcal B_z(x)
 =\frac{2\varrho(x)}{1+\sqrt{1-\lambda^2}}T_z(x)
 =\frac{2p(x)}{1+\sqrt{1-\lambda^2}}LA_t(x)L^{-1},
 \qquad z=e^{it}.
\end{align}
The scalar factor is positive and bounded away from zero.
It cancels from the ratios defining dominated splitting, and
constant conjugation preserves that property. Since $A_t$ has
determinant one, dominated splitting is equivalent to uniform
hyperbolicity. Therefore \cite[Theorem~4.11]{CFO23} gives
\begin{align}\label{eq:boundary-spectral}
 e^{it}\notin\Sigma_{1,\lambda,\alpha}
 \quad\Longleftrightarrow\quad
 (\alpha,A_t)\text{ is uniformly hyperbolic}.
\end{align}
This justifies the spectral criterion used in both propositions
below without invoking a singular real cocycle at $\lambda=1$.

\subsection{No gaps with even winding number}
\begin{proposition}[$L^*$]\label{prop:boundary-even-closure}
Let $\alpha$ be irrational and let $0<\lambda<1$.
No gap with even winding number is open in
$\Sigma_{1,\lambda,\alpha}$ or $\Sigma_{\lambda,1,\alpha}$.
\end{proposition}

\begin{proof}
We first consider $(\lambda_1,\lambda_2)=(1,\lambda)$.
The formula \eqref{eq:real-cocycle} remains well-defined on
this edge: here $c=\lambda_1'=0$ and
\begin{align}
 p(x)=\sqrt{1-\lambda^2\sin^2(2\pi x)}>0.
\end{align}
Since $\sin(2\pi(x+\tfrac12))=-\sin(2\pi x)$,
direct substitution gives
\begin{align}\label{eq:boundary-half-period}
 A_t\left(x+\tfrac12\right)=J A_t(x)J^{-1},
 \qquad
 J=\begin{pmatrix}0&-1\\1&0\end{pmatrix}.
\end{align}

Suppose that $e^{it}$ belongs to an open spectral gap.
By \eqref{eq:boundary-spectral}, the normalized real cocycle
$(\alpha,A_t)$ is uniformly hyperbolic. Thus it has a unique
continuous stable section
$s:\T\to\mathbb{RP}^1$.
The identity \eqref{eq:boundary-half-period} holds for every
iterated product, so uniqueness of the stable section \cite[Section 2.1]{Avila2015Acta} implies
\begin{align}
 s\left(x+\tfrac12\right)=J s(x).
\end{align}

Choose a continuous angle lift $\vartheta:\R\to\R$ such that
\begin{align}
 s(x)=
 \operatorname{span}
 \begin{pmatrix}\cos\vartheta(x)\\ \sin\vartheta(x)\end{pmatrix}.
\end{align}
The matrix $J$ rotates through $\pi/2$, while two angles
describe the same line precisely when they differ by an
integer multiple of $\pi$. Therefore
\begin{align}
 \vartheta\left(x+\tfrac12\right)-\vartheta(x)
 =\frac{\pi}{2}+q(x)\pi,
 \qquad q(x)\in\Z.
\end{align}
Moreover,
\begin{align}
 q(x)=\frac{
 \vartheta(x+\tfrac12)-\vartheta(x)-\pi/2}{\pi}
\end{align}
is continuous. A continuous integer-valued function on $\R$
is constant, by the intermediate value theorem. Hence
$q(x)=q$ for one fixed integer $q$.

Applying the same identity twice gives
\begin{align}
 \vartheta(x+1)-\vartheta(x)
 &=
 \bigl[\vartheta(x+1)-\vartheta(x+\tfrac12)\bigr]
 +\bigl[\vartheta(x+\tfrac12)-\vartheta(x)\bigr] =2\left(\frac{\pi}{2}+q\pi\right)
 =(2q+1)\pi.
\end{align}
Consequently the stable line has odd projective degree:
\begin{align}
 \deg(s)=\frac{\vartheta(x+1)-\vartheta(x)}{\pi}=2q+1.
\end{align}

This degree is the winding integer in \eqref{eq:labels}.
Indeed, choose a continuous projective frame whose first
column spans $s(x)$. Its degree is $\deg(s)$, and the
conjugated cocycle fixes a constant line, so its rotation
number belongs to $\frac12\Z$. The conjugacy-degree formula
\cite[p.~102, equation~(2.9)]{AJ2012JEMS} therefore yields
\begin{align}
 2\rho(\alpha,A_t)\equiv\deg(s)\alpha\pmod1.
\end{align}
Thus the winding of every open gap is odd.

Finally, coupling-exchange duality preserves the spectrum
and its labels
\cite[Corollary~2.6]{CFO23},
\cite[Theorem~4.3]{CL26}, with the boundary extension by continuity explained at the end of the proof of Proposition~\ref{prop:odd-gap-seeds}.
The same conclusion follows on $(\lambda,1)$.
\end{proof}

\subsection{Seeds for Odd Gaps}\label{subsec:boundary}
Having ruled out even gaps, our main goal of this section is to provide odd seed gaps that we can extend to the entire boundary with the tools we will develop in Section~\ref{Sec:montonicity}.

\begin{proposition}[$L^*$]\label{prop:odd-gap-seeds}
For every irrational $\alpha$ and every odd integer $m$, there is
$\eps_{\alpha,m}\in(0,1)$ such that all gaps with winding number
$m$ are open in $\Sigma_{1,\lambda,\alpha}$ and
$\Sigma_{\lambda,1,\alpha}$ whenever $0<\lambda<\eps_{\alpha,m}$.
Equivalently, each value $\rho_*$ satisfying
$2\rho_*\equiv m\alpha\pmod1$ occurs on an open gap.
\end{proposition}

For positive odd $m$, the proof proceeds in three steps.
Lemma~\ref{lem:boundary-resonant-profile} constructs an approximate
Fourier solution whose leading residual is a nonzero resonant term
of order $\lambda^m$.
Lemma~\ref{lem:boundary-resonant-splitting} converts this residual
into a constant symmetric traceless term in a real normal form,
while keeping track of the degree of the conjugating frame.
Lemma~\ref{lem:symmetric-seed-splitting} then shows that this term
dominates the remainder and produces uniform hyperbolicity with
the required rotation number, yielding the desired seed gap.
This final criterion will also be used for the interior seeds
in Appendix~\ref{app:interior-seeds}.

\begin{lemma}[$L^*$]\label{lem:boundary-resonant-profile}
Fix an irrational $\alpha$, a positive odd integer $m$, and
$t_0\in\mathbb R$ such that $e^{2it_0}=e^{2\pi im\alpha}$.
Set $z_0=e^{it_0}$ and let
\begin{align}
 M_{t,\lambda}(x)=
 \begin{pmatrix}e^{it}&-\lambda\sin(2\pi x)\\
 -\lambda\sin(2\pi x)&e^{-it}\end{pmatrix}.
\end{align}
For sufficiently small real $\lambda$, there are real-valued analytic
functions $t(\lambda),s(\lambda)$ and a vector-valued
trigonometric polynomial $v=v_\lambda$, with coefficients real
analytic in $\lambda$, such that
\begin{align}
 t=t_0+O(\lambda^2),\qquad s=1+O(\lambda^2),\qquad
 v=e_1+O(\lambda),
\end{align}
and
\begin{align}\label{eq:odd-seed-residual}
 M_{t(\lambda),\lambda}v-z_0s(\lambda)v(\cdot+\alpha)
 =c_m\lambda^m e^{-2\pi imx}e_2+O(\lambda^{m+1})
\end{align}
for a constant $c_m\ne0$. The errors are uniform in $x$.
\end{lemma}

\begin{proof}
Let $q=e^{2\pi i\alpha}$ and $\kappa=1/(2i)$, and recall $e_1$ and $e_2$ are the standard basis of $\R^2$.
Since $z_0^2=q^m$, both $e_1$ and $e^{-2\pi imx}e_2$ solve
$M_{t_0,0}v=z_0v(\cdot+\alpha)$. We construct an approximate
solution near $e_1$, leaving a residual in the second resonant mode.
All constants below may depend on $\alpha,m,t_0$.

\smallskip\noindent\emph{Finite cancellation.}
The off-diagonal part of $M_{t,\lambda}(x)$ sends a Fourier mode
of index $k$ in one component to a linear combination of modes
of indices $k-1$ and $k+1$ in the other component. Thus each shift changes both the
parity of the Fourier index and the component. Accordingly, let
$e(k)=e_1$ for even $k$ and $e(k)=e_2$ for odd $k$, and seek
\begin{align}
 v(x)=\sum_{k\in I}a_ke^{2\pi ikx}e(k),\qquad
 I=\{-m+1,\ldots,m\},\qquad a_0=1.
\end{align}
Set $a_k=0$ outside $I$. The first omitted negative index is the
resonance $-m$, whereas the first omitted positive index is $m+1$.
The latter is one step farther from zero and will contribute only
at a higher order in $\lambda$.

Introduce an auxiliary real scalar $s$, initially equal to one.
Since $-\sin(2\pi x)=\kappa(e^{-2\pi ix}-e^{2\pi ix})$,
the coefficient at index $k$ of
$M_{t,\lambda}v-z_0s v(\cdot+\alpha)$ is $F_ke(k)$, where
\begin{align}\label{eq:odd-seed-finite-system}
 F_k=f_k(t,s)a_k+\kappa\lambda(a_{k+1}-a_{k-1}),
 \qquad f_k(t,s)=e^{i(-1)^kt}-z_0s q^k.
\end{align}
We solve $F_k=0$ for all $k\in I$, using the real unknowns $t,s$
and the complex unknowns $a_k$, $k\in I\setminus\{0\}$.
The scalar $s$ supplies the amplitude adjustment needed, together
with the phase adjustment $t$, to solve the complex equation at zero.

At $\lambda=0$, the solution is $(t,s,a)=(t_0,1,\delta_0)$.
Writing $d_k=f_k(t_0,1)$ gives
\begin{align}
 d_k=
 \begin{cases}
 z_0(1-q^k),&k\text{ even},\\
 z_0^{-1}(1-q^{m+k}),&k\text{ odd}.
 \end{cases}
\end{align}
Since $\alpha$ is irrational, $d_k$ vanishes only at $k=0$
and $k=-m$. Thus $d_k\ne0$ for $k\in I\setminus\{0\}$.
The derivative of the finite system in its unknowns is
\begin{align}
 \delta F_0=z_0(i\,\delta t-\delta s),\qquad
 \delta F_k=d_k\,\delta a_k\quad(k\ne0).
\end{align}
This is invertible over $\R$, so the real analytic implicit function
theorem gives $t(\lambda),s(\lambda),a_k(\lambda)$ solving the
system. In particular, $a_k=O(\lambda)$ for $k\ne0$.
The equation at zero then yields
\begin{align}
 e^{it}-z_0s=-\kappa\lambda(a_1-a_{-1})=O(\lambda^2),
 \qquad t=t_0+O(\lambda^2),\quad s=1+O(\lambda^2)>0.
\end{align}

\smallskip\noindent\emph{The surviving coefficient.}
For $k\in I\setminus\{0\}$, equation
\eqref{eq:odd-seed-finite-system} reads
\begin{align}\label{eq:odd-seed-recurrence}
 a_k=-\frac{\kappa\lambda}{f_k}(a_{k+1}-a_{k-1}),
\end{align}
and its denominator stays bounded away from zero.
This implies $a_k=O(\lambda^{|k|})$: starting with the bound
$O(\lambda)$ away from zero, if $a_k=O(\lambda^j)$ for $|k|\ge j$,
then both neighbors of an index with $|k|\ge j+1$ have that bound,
and \eqref{eq:odd-seed-recurrence} supplies another factor of
$\lambda$. The zero coefficients outside $I$ satisfy every such bound.

For $1\le j\le m-1$, we therefore have
$f_{-j}=d_{-j}+O(\lambda^2)$ and $a_{-j-1}=O(\lambda^{j+1})$.
Substitution into \eqref{eq:odd-seed-recurrence} gives
\begin{align}
 a_{-j}=-\frac{\kappa\lambda}{d_{-j}}a_{-j+1}
             +O(\lambda^{j+2}).
\end{align}
Iterating from $a_0=1$ gives
\begin{align}
 a_{-j}=\frac{(-\kappa)^j\lambda^j}{d_{-1}\cdots d_{-j}}
             +O(\lambda^{j+2}).
\end{align}
All residual coefficients in $I$ vanish exactly. The only remaining
ones are at $-m$ and $m+1$, so
\begin{align}
 M_{t(\lambda),\lambda}v-z_0s(\lambda)v(\cdot+\alpha)
 &=\kappa\lambda a_{-m+1}e^{-2\pi imx}e_2
   -\underbrace{\kappa\lambda a_m e^{2\pi i(m+1)x}e_1}_{O(\lambda^{m+1})},
\end{align}
where, since $m-1$ is even,
\begin{align}
 c_m=\frac{\kappa^m}{d_{-1}\cdots d_{-(m-1)}}\ne0.
\end{align}
For $m=1$, the product is empty and $a_{-m+1}=a_0=1$, so the
same formula applies.
\end{proof}

It remains to convert this complex residual into a symmetric
splitting of the real cocycle.

\begin{lemma}[$L^*$]\label{lem:boundary-resonant-splitting}
Fix an irrational $\alpha$, a positive odd integer $m$, and
$t_0\in\mathbb R$ such that $e^{2it_0}=e^{2\pi im\alpha}$.
Set $\sigma=e^{it_0-\pi im\alpha}\in\{1,-1\}$, and write
$A_{t,\lambda}$ for the real cocycle on the edge
$(\lambda_1,\lambda_2)=(1,\lambda)$.
For sufficiently small positive $\lambda$, there are a real analytic
$t(\lambda)=t_0+O(\lambda^2)$, a real analytic determinant-one projective frame
$B_\lambda(x)=R_{\pi mx}+O(\lambda)$ of degree $m$, and a nonzero
constant real symmetric traceless matrix $H$ such that
\begin{align}\label{eq:boundary-resonant-splitting}
 \sigma B_\lambda(x+\alpha)^{-1}
 A_{t(\lambda),\lambda}(x)B_\lambda(x)
 =I+\lambda^mH+O(\lambda^{m+1}).
\end{align}
Here $R_\vartheta=e^{\vartheta J}$, and the error is uniform in $x$.
\end{lemma}

\begin{proof}
Take $t,s,v,c_m$ from Lemma~\ref{lem:boundary-resonant-profile},
retaining its notation $z_0$ and $M_{t,\lambda}$. On the edge
$(1,\lambda)$ we have $\mathcal S_{e^{it}}=p_\lambda^{-1}M_{t,\lambda}$,
where $p_\lambda(x)=\sqrt{1-\lambda^2\sin^2(2\pi x)}$.

\smallskip\noindent\emph{The projective frame.}
Use $v=e_1+O(\lambda)$ to form
\begin{align}
 C(x)=\begin{pmatrix}
 e^{\pi imx}v_1(x)&e^{-\pi imx}\overline{v_2(x)}\\
 e^{\pi imx}v_2(x)&e^{-\pi imx}\overline{v_1(x)}
 \end{pmatrix}
 \end{align}\
with
\begin{align}
 \Delta(x)=\det C(x)=|v_1(x)|^2-|v_2(x)|^2=1+O(\lambda)>0.
\end{align}
Denote $C_+=C(x+\alpha)$ and $\zeta=c_me^{\pi im\alpha}$,
and retain the sign $\sigma=z_0e^{-\pi im\alpha}$ fixed above.
Multiplying \eqref{eq:odd-seed-residual} by $e^{\pi imx}$
gives the scalar multiplier $z_0s e^{-\pi im\alpha}=\sigma s$
for the first column. After applying $C_+^{-1}$, its lower residual
entry is
\begin{align}
 c_m\lambda^me^{\pi im\alpha}
 \frac{v_1(x+\alpha)}{\Delta(x+\alpha)}
 +O(\lambda^{m+1})
 =\zeta\lambda^m+O(\lambda^{m+1}),
\end{align}
while the upper residual entry has the additional factor
$\overline{v_2(x+\alpha)}=O(\lambda)$.
The antilinear map $(w_1,w_2)\mapsto(\overline{w_2},\overline{w_1})$
commutes with $M_{t,\lambda}$ and exchanges the columns of $C$.
It therefore supplies the second column, giving
\begin{align}
 N(x):=C_+^{-1}M_{t(\lambda),\lambda}C
 =\sigma sI+\lambda^mH_\zeta+O(\lambda^{m+1}),\qquad
 H_\zeta=\begin{pmatrix}0&\overline\zeta\\\zeta&0\end{pmatrix}.
\end{align}

\smallskip\noindent\emph{Determinant normalization and degree.}
The normalized real frame
$B=\Delta^{-1/2}U_C^{-1}CU_C\in\mathrm{SL}(2,\R)$ satisfies
\begin{align}
 B(x+\alpha)^{-1}A_{t(\lambda),\lambda}(x)B(x)
 =\omega(x)U_C^{-1}N(x)U_C,\qquad
 \omega(x)=\frac{\sqrt{\Delta(x+\alpha)/\Delta(x)}}{p_\lambda(x)}>0.
\end{align}
Since the left side has determinant one and
$\det N=s^2+O(\lambda^{m+1})$, we obtain
$\omega s=1+O(\lambda^{m+1})$.
Here we used $\operatorname{tr}H_\zeta=0$ and $2m\ge m+1$.
Also $s=1+O(\lambda^2)$, so
\begin{align}
 \sigma B(x+\alpha)^{-1}A_{t(\lambda),\lambda}(x)B(x)
 =I+\lambda^m\sigma U_C^{-1}H_\zeta U_C+O(\lambda^{m+1}).
\end{align}
The real matrix
\begin{align}
 U_C^{-1}H_\zeta U_C
 =\begin{pmatrix}
 \operatorname{Re}\zeta&-\operatorname{Im}\zeta\\
 -\operatorname{Im}\zeta&-\operatorname{Re}\zeta
 \end{pmatrix}
\end{align}
is symmetric with eigenvalues $\pm\beta$, where $\beta=|\zeta|>0$.
Set $H=\sigma U_C^{-1}H_\zeta U_C$ and $B_\lambda=B$.
This gives \eqref{eq:boundary-resonant-splitting} with $H\ne0$.
Since $C(x+1)=(-1)^mC(x)$, the normalized frame is projectively
periodic. At $\lambda=0$ its real form is $R_{\pi mx}$, and
$B_\lambda=R_{\pi mx}+O(\lambda)$; hence its projective degree is $m$.
\end{proof}

\begin{lemma}[$L^*$]\label{lem:symmetric-seed-splitting}
Fix an irrational $\alpha$, an integer $m$, and $\sigma\in\{1,-1\}$.
Let $A_\lambda:\mathbb T\to\mathrm{SL}(2,\mathbb R)$ be continuous
cocycles homotopic to a constant. Suppose that there are continuous
frames $B_\lambda:\mathbb R\to\mathrm{SL}(2,\mathbb R)$ of projective
degree $m$, with $B_\lambda(x+1)=(-1)^mB_\lambda(x)$, such that
\begin{align}\label{eq:shared-seed-splitting}
 \sigma B_\lambda(x+\alpha)^{-1}A_\lambda(x)B_\lambda(x)
 =I+D_\lambda+E_\lambda(x).
\end{align}
Assume that $D_\lambda$ is a constant real symmetric traceless matrix,
that $d_\lambda:=\|D_\lambda\|>0$ tends to zero as $\lambda\downarrow0$,
and that $\|E_\lambda\|_\infty=o(d_\lambda)$.
Then, for all sufficiently small positive $\lambda$,
$(\alpha,A_\lambda)$ is uniformly hyperbolic and
\begin{align}
 \rho(\alpha,A_\lambda)
 =\frac{m\alpha}{2}+\frac{1-\sigma}{4}\pmod1.
\end{align}
\end{lemma}

\begin{proof}
A constant rotation takes $D_\lambda$ to
$d_\lambda\left(\begin{smallmatrix}0&1\\1&0\end{smallmatrix}\right)$.
The normalized cocycle then has determinant one and the form
\begin{align}
 F_\lambda(x)=
 \begin{pmatrix}1&d_\lambda\\d_\lambda&1\end{pmatrix}
 +o(d_\lambda).
\end{align}
For small positive $\lambda$, all entries are positive and every
row and column sum is at least $1+d_\lambda/2$.
It maps the positive quadrant strictly into itself and expands the
$\ell^1$ norm there by this factor. If
\begin{align}
    F_\lambda= \begin{pmatrix}a&b\\c&d\end{pmatrix}\quad\text{and}\quad H_{0} = \begin{pmatrix}
        1 & 0 \\ 0 & -1
    \end{pmatrix},
\end{align}
then
\begin{align}
 H_0F_\lambda^{-1}H_0=
 \begin{pmatrix}d&b\\c&a\end{pmatrix}.
\end{align}
The same bounds give a backward expanding cone in the
opposite-sign quadrant. The forward and backward cone criterion
therefore gives uniform hyperbolicity.

Choose the angle lift of $F_\lambda$ near the identity. Forward
orbits in the positive quadrant stay in an interval of length
$\pi/2$, so $\rho(\alpha,F_\lambda)=0$ modulo one.
Restoring $\sigma$ adds $(1-\sigma)/4$. The constant rotation does
not change the degree of the frame, and the conjugacy-degree formula
\cite[p.~102, equation~(2.9)]{AJ2012JEMS} adds $m\alpha/2$.
Bounded conjugation and multiplication by $\sigma$ preserve uniform
hyperbolicity, proving both conclusions.
\end{proof}

\begin{proof}[Proof of Proposition~\ref{prop:odd-gap-seeds}]
Fix a positive odd $m$ and either permitted value $\rho_*$.
Choose $t_0$ with $t_0/(2\pi)=\rho_*$ modulo one. Then
$e^{2it_0}=e^{2\pi im\alpha}$ and
$\sigma=e^{it_0-\pi im\alpha}\in\{1,-1\}$.
Lemma~\ref{lem:boundary-resonant-splitting} gives a degree-$m$ frame
and the splitting $D_\lambda=\lambda^mH$, with
$d_\lambda=\|H\|\lambda^m>0$ and remainder
$O(\lambda^{m+1})=o(d_\lambda)$.
Lemma~\ref{lem:symmetric-seed-splitting} therefore gives uniform
hyperbolicity and
\begin{align}
 \rho\bigl(\alpha,A_{t(\lambda),\lambda}\bigr)
 =\frac{m\alpha}{2}+\frac{1-\sigma}{4}
 =\rho_*\pmod1.
\end{align}
By \eqref{eq:boundary-spectral}, $e^{it(\lambda)}$ is in the resolvent.
Uniform hyperbolicity persists on an open interval of $t$, with the
same rotation-number value, so the corresponding gap is open.

Finally, $A_{-t}=H_0A_tH_0^{-1}$ with
$H_0=\operatorname{diag}(1,-1)$ reverses rotation number and supplies
the negative odd windings. Since duality exchanges the couplings and preserves the
spectrum and its gap labels
\cite[Corollary~2.6]{CFO23}, \cite[Theorem~4.3 and its proof]{CL26},
equality of the density-of-states measures extends to the boundary
by norm continuity. Thus the same labels are open on $(\lambda,1)$.
Taking the smaller threshold for the two permitted values proves
the proposition.
\end{proof}

\section{Monotonicity of the height function}~\label{Sec:montonicity}

\begin{definition}[Magnetic rotation algebra and canonical trace]
For the irrational frequency $\alpha$ of Section~\ref{Sec:Intro}, let $\mathcal A$
be the universal unital $C^*$-algebra generated by unitaries $u,v$ with
\begin{align}
 uv&=e^{-2\pi i\alpha}vu,\qquad
 \tr(u^jv^k)=\delta_{j0}\delta_{k0},
\end{align}
where $\tr$ is the faithful normalized canonical trace.
In the convention of~\eqref{eq:model}, $(u\psi)_j=\psi_{j-1}$ and
$(v\psi)_j=e^{2\pi i(j\alpha+\theta)}\psi_j$.
\end{definition}

We regard $W=W_{\lambda_1,\lambda_2}$ from~\eqref{eq:model} as a
$2\times2$ matrix over $\mathcal A$ and reduce its real part to a
scalar operator whose spectrum determines $\sigma(W)$. This scalar
representation puts the operator in the form used by the height-function
monotonicity argument of \cite{BorgniaFaust2026} for the critical
almost Mathieu operator.

\subsection{Scalar Magnetic Representation}\label{subsec:flatten}
\begin{definition}[Coupling parameters]
By coupling-exchange duality, we work with
$0<\lambda _2\leq\lambda _1\leq1$. Retain $r$ from~\eqref{eq:phasefactor},
now allowing $r=1$, and set
\begin{align}
 \label{eq:half-angle}
 a&=\frac{\lambda _1}{1+c},\quad
 \lambda _1=\frac{2a}{1+a^2},\quad
 \lambda _2=\frac{2r}{1+r^2},\quad
 M=(1+a^2)(1+r^2).
\end{align}
Thus $0<r\leq a\leq1$; all radial derivatives and gap continuations
hold $a$ fixed.
\end{definition}

\smallskip\noindent\emph{Trace representation.}
Represent $\mathcal A$ by left multiplication on
\begin{align}
 \mathscr H&=L^2(\mathcal A,\tr),\qquad
 \langle A,B\rangle=\tr(A^*B),\qquad \Omega=I.
\end{align}
For the physical space $\mathcal H$ of Section~\ref{Sec:Intro}, take the phase
direct integral $L^2(\mathbb T;\mathcal H)$. Fourier transformation
in $\theta$ sends $\psi$ to
\begin{align}
 X^{\pm}_{j,k}
 &=\int_0^1e^{-2\pi ik\theta}\psi_j^{\pm}(\theta)\,d\theta,
 \qquad
 \mathcal F_\theta\psi
 =\sum_{j,k}\begin{pmatrix}X^+_{j,k}\\X^-_{j,k}\end{pmatrix}
 u^jv^k\Omega.
\end{align}
The vectors $u^jv^k\Omega$ form an orthonormal basis of $\mathscr H$,
and Parseval gives
\begin{align}
 \int_0^1\sum_{j\in\mathbb Z}
 \bigl(|\psi_j^+(\theta)|^2+|\psi_j^-(\theta)|^2\bigr)\,d\theta
 &=\sum_{j,k\in\mathbb Z}
 \bigl(|X^+_{j,k}|^2+|X^-_{j,k}|^2\bigr)
 =\|\mathcal F_\theta\psi\|^2.
\end{align}
Surjectivity follows from inverse Fourier transformation, so
$\mathcal F_\theta$ is unitary onto $\mathscr H\oplus\mathscr H$.
On each spin component, the generators act on Fourier coefficients by
\begin{align}
 (uX^{\pm})_{j,k}&=X^{\pm}_{j-1,k},\qquad
 (vX^{\pm})_{j,k}=e^{2\pi ij\alpha}X^{\pm}_{j,k-1}.
\end{align}
These are precisely left multiplication by $u$ and $v$ in the basis
$u^jv^k\Omega$, since $vu^j=e^{2\pi ij\alpha}u^jv$.
In this representation the UAMO is
\begin{align}
 \label{eq:shifted-walk}
 W&=\overbrace{\frac1{1+a^2}
 \begin{pmatrix}2au&a^2-1\\1-a^2&2au^*\end{pmatrix}}^{S_{\lambda_1}}
 \overbrace{\frac1{1+r^2}
 \begin{pmatrix}
 r(v+v^*)+i(1-r^2)I&ir(v-v^*)\\
 -ir(v-v^*)&r(v+v^*)-i(1-r^2)I
 \end{pmatrix}}^{Q_{\lambda_2}}.
\end{align}
The unitary $\mathcal F_\theta$ therefore intertwines the phase family
with $W$ on $\mathscr H\oplus\mathscr H$. Faithfulness and phase
independence from Section~\ref{Sec:Intro} give
\begin{align}
 \label{eq:walk-algebra-spectrum}
 \sigma_{M_2(\mathcal A)}(W)
 &=\sigma(W_{\lambda _1,\lambda _2,\alpha,\theta})
 =\Sigma_{\lambda _1,\lambda _2,\alpha}.
\end{align}
The normalized matrix trace $\tr_2$ averages the scalar traces of
the two diagonal entries. In the density-of-states normalization of
Sections~\ref{Sec:Intro}--\ref{Sec:one}, for $f\in C(\partial\mathbb D)$,
\begin{align}
 \tr_2(f(W))
 &:=\frac12\sum_{s\in\{+,-\}}\int_0^1
 \langle\delta_0^s,f(W_{\lambda _1,\lambda _2,\alpha,\theta})
 \delta_0^s\rangle\,d\theta
 =\int_{\partial\mathbb D}f(z)\,d\nu_r(z).
\end{align}
Here $\nu_r$ is the unitary density-of-states measure. The phase
average extracts the identity coefficient: on monomials,
$u^jv^k\delta_0=e^{2\pi ik\theta}\delta_j$; norm density then gives
the equality for $f(W)$.

\smallskip\noindent\emph{Reflections and scalar blocks.}
With the spin flip $X_0$ from~\eqref{eq:spin-flip}, set
\begin{align}
 R_1&=S_{\lambda_1}X_0,\qquad R_2=X_0Q_{\lambda_2},
 \qquad \sigma_z=\operatorname{diag}(1,-1).
\end{align}

\begin{lemma}[Reflection symmetries; $L^*$]
\label{lem:reflection}\label{lem:spin-blocks}
The operators $R_1,R_2$ are self-adjoint unitaries. The unitary
\begin{align}
 B&=\frac1{\sqrt{1+a^2}}\begin{pmatrix}au&-u\\I&aI\end{pmatrix}
\end{align}
diagonalizes $R_1$:
\begin{align}
 \label{eq:reflection-walk-blocks}
 B^*R_1B&=\sigma_z\otimes I,\quad
 W=R_1R_2,\quad R_1WR_1=W^*,\quad
 B^*WB=(\sigma_z\otimes I)B^*R_2B.
\end{align}
Both $\sigma(W)$ and $\nu_r$ are invariant under conjugation and
the antipodal map.
\end{lemma}

\begin{proof}
The identities $X_0S_{\lambda_1}X_0=S_{\lambda_1}^*$ and
$X_0Q_{\lambda_2}X_0=Q_{\lambda_2}^*$ give $R_j^*=R_j$ and
$R_j^2=I$, hence the identities for $W$. The $+1$ and $-1$ eigenspaces
of $R_1$ are parametrized by
$(auf,f)$ and $(-uf,af)$. These parametrizations are orthogonal and
multiply norms by $\sqrt{1+a^2}$, so the columns of $B$ give the
claimed unitary diagonalization.

Since $W$ and $W^*$ are unitarily conjugate, their spectra agree;
invariance of $\tr_2$ also gives conjugation symmetry of $\nu_r$.
In the physical basis,
\begin{align}
 W_{\lambda_1,\lambda_2,\alpha,\theta+1/2}
 &=-\overline{W_{\lambda_1,\lambda_2,\alpha,\theta}},
\end{align}
since the shift is real and the phase shift replaces $v$ by $-v$.
Phase independence gives the spectral symmetry $z\mapsto-\overline z$,
and phase averaging gives the same symmetry for $\nu_r$.
Combining it with conjugation gives the antipodal symmetry.
\end{proof}

\begin{lemma}[Spectral correspondence; $L^*$]\label{lem:cosine}
\begin{align}
 \label{eq:cosine-spectrum}
 e^{it}\in\Sigma_{\lambda_1,\lambda_2,\alpha}
 &\quad\Longleftrightarrow\quad\cos t\in\sigma(\operatorname{Re}W).
\end{align}
Each bounded real gap lifts to two conjugate unitary gaps, and the
real gap closes exactly when its lifts close.
\end{lemma}

\begin{proof}
Spectral mapping gives
$\sigma(\operatorname{Re}W)=\{\operatorname{Re}z:z\in\sigma(W)\}$.
The two circle points with a given real part are conjugate.
By Lemma~\ref{lem:reflection}, either both belong to $\sigma(W)$ or
neither does; at $\pm1$ they coincide. On either open semicircle,
cosine is a homeomorphism, so gaps and their endpoints correspond.
\end{proof}

Taking the real part of \eqref{eq:reflection-walk-blocks} cancels
the off-diagonal blocks because $B^*R_2B$ is self-adjoint:
\begin{align}
 \label{eq:real-walk-blocks}
 B^*\operatorname{Re}WB
 &=\begin{pmatrix}(B^*R_2B)_{11}&0\\0&-(B^*R_2B)_{22}\end{pmatrix}.
\end{align}
The two diagonal blocks are related by magnetic inversion.

\begin{definition}[Magnetic inversion]
Define a complex-linear $*$-automorphism on the generators and
its implementing operator on trace vectors by
\begin{align}
 \iota(u)&=-u^*,\qquad \iota(v)=-e^{-2\pi i\alpha}v^*,\qquad
 \mathcal C(F\Omega)=\iota(F)\Omega.
\end{align}
These assignments preserve the magnetic relation and canonical trace,
and $\iota^2=I$. Thus $\iota$ is an involution, implemented
by $\mathcal C=\mathcal C^*=\mathcal C^{-1}$, since
$\|\iota(F)\Omega\|^2=\tr(\iota(F^*F))=\tr(F^*F)$.
\end{definition}
\begin{lemma}[The scalar blocks; $L^*$]\label{lem:block-symmetry}
Setting $d:=1-a^2e^{2\pi i\alpha}$, define
\begin{align}
 \label{eq:intermediate-harper}
 K_r&:=\frac{2a}{M}\operatorname{Re}\!\bigl[i(I-irv^*)(I-irv)u\bigr]
       +\frac{ir}{M}(dv-\bar d v^*).
\end{align}
Then,
\begin{align}
 B^*\operatorname{Re}WB&=K_r\oplus\bar K_r,\qquad
 \bar K_r:=\mathcal C^*K_r\mathcal C,\qquad
 \sigma(\operatorname{Re}W)=\sigma(K_r).
\end{align}
\end{lemma}

\begin{proof}
Pair the first column of $B$ with $R_2=X_0Q_{\lambda_2}$ and use
$u^*vu=e^{2\pi i\alpha}v$:
\begin{align}
 (B^*R_2B)_{11}
 &=\frac{2a}{1+a^2}\operatorname{Re}\!\left[
 \left(\frac{\lambda_2}{2}(v+v^*)+i\lambda_2'I\right)u\right]
 +\frac{i\lambda_2}{2(1+a^2)}(dv-\bar d v^*).
\end{align}
The half-angle formulas and
$i(I-irv^*)(I-irv)=r(v+v^*)+i(1-r^2)I$
identify this with $K_r$ in \eqref{eq:intermediate-harper}.

The signed blocks have the same hopping term. It is fixed by $\iota$,
which sends $i(I-irv^*)(I-irv)u$ to its adjoint. Computing the second
column and comparing the potential terms gives
\begin{align}
 -(B^*R_2B)_{22}-K_r
 &=\frac{ir}{1+r^2}\bigl((e^{2\pi i\alpha}-1)v
                         -(e^{-2\pi i\alpha}-1)v^*\bigr)
 =\iota(K_r)-K_r.
\end{align}
Substitution into \eqref{eq:real-walk-blocks} gives the direct sum;
unitary equivalence of the blocks gives the spectral equality.
\end{proof}

\smallskip\noindent\emph{Gauge normalization.}
To separate the radial powers in the hopping, fix $v$ and replace
$u$ by $\omega_r(v)u$, where
\begin{align}
 \omega_r(v)&=
 \begin{cases}
 (I+irv)(I-irv^*)^{-1},&r<1,\\
 iv,&r=1,
 \end{cases}
 \qquad
 G_r(u^jv^k\Omega)=(\omega_r(v)u)^jv^k\Omega.
\end{align}
For $r<1$, the commuting numerator and denominator have equal
modulus, and the denominator is invertible; hence $\omega_r(v)$
is unitary. At $r=1$ we define it separately as the unitary $iv$.
The substitution preserves the magnetic relation and has inverse
$u\mapsto\omega_r(v)^*u$, $v\mapsto v$. It also preserves the trace:
nonzero $u$-degree stays nonzero, and powers of $v$ are fixed.
Thus $G_r$ is unitary, and the hopping simplifies
\begin{align}
 (I-irv^*)(I-irv)\omega_r(v)&=I+r^2v^2.
\end{align}
This identity also holds at $r=1$ by direct multiplication.
Introduce the fixed generators
\begin{align}
 U&=-vu,\qquad V=iv,\qquad UV=e^{-2\pi i\alpha}VU.
\end{align}
Since the gauge fixes the potential, the normalized operator
$H_r:=(M/a)G_rK_rG_r^*$ takes the form
\begin{align}
 \label{eq:harper}
H_r=(V^*-r^2V)U+U^*(V-r^2V^*)
 +r\bigl((a^{-1}-ae^{2\pi i\alpha})V+(a^{-1}-ae^{-2\pi i\alpha})V^*\bigr).
\end{align}
The unitary gauge and Lemmas~\ref{lem:cosine} and~\ref{lem:block-symmetry} give
\begin{align}
 \label{eq:harper-spectrum}
 e^{it}\in\Sigma_{\lambda_1,\lambda_2,\alpha}
 &\quad\Longleftrightarrow\quad (M/a)\cos t\in\sigma(H_r).
\end{align}
Subsequent radial derivatives act on the displayed polynomial
$H_r$, with $U,V$ fixed. This polynomial is norm-continuous up to
$r=1$. No norm limit of the gauges $G_r$ is required: the gauge
identities hold separately at each $r$, including $r=1$.

\subsection{The height function maximum}
Following~\cite{BorgniaFaust2026}, we compare the logarithmic potential
of $H_r$ with the UAMO Lyapunov exponent and track its maximum in each gap.

\begin{definition}[Scalar density of states]\label{def:scalar-dos}
Let $\mu_r$ be the trace spectral measure of $H_r$, characterized by
$\int f(x)\,d\mu_r(x)=\tr f(H_r)$ for continuous $f$. And, its real integrated density of states is
\begin{align}
N_r(E):=\mu_r(( -\infty,E]).
\end{align}
\end{definition}

\begin{definition}[Internal gaps and gap branches]\label{def:gap-branches}
An \emph{internal gap} $G_n(r)$ is a bounded connected component of
$\mathbb R\setminus\sigma(H_r)$.
A \emph{gap branch} is a family $G_n(r)$ on a parameter interval
such that, near each parameter, one fixed open energy interval lies
in all the chosen gaps. The index $n$ distinguishes real gap branches;
it is not the winding $m$ of a unitary gap in~\eqref{eq:labels}.
\end{definition}

\begin{definition}[Height functions]\label{def:heights}
For $z\in\partial\mathbb D$, define the normalized unitary height by
\begin{align}
 \label{eq:unitary-height}
 h_r^W(z)&:=2\tr_2\bigl(\log|z-W|\bigr)+\log\frac M{2a}.
\end{align}
For real $E$, define the real height and its supremum over an
internal gap by
\begin{align}
 \label{eq:height-definition}
 h_r(E)&:=\tr\bigl(\log|E-H_r|\bigr),\qquad
 h_n(r):=\sup_{E\in G_n(r)}h_r(E).
\end{align}
The logarithmic trace is $\int\log|E-x|\,d\mu_r(x)$, interpreted by
decreasing truncations $\max\{\log|E-x|,-K\}$ as $K\to\infty$.
Use the same convention for the unitary trace with $\nu_r$ from
Section~\ref{subsec:flatten}.
Partial differentiation in $r$ holds $E$ fixed.
\end{definition}

\begin{lemma}[$L^*$]\label{lem:height-label-properties}
The heights $h_r$ and $h_r^W$ are finite off their respective spectra.
The function $N_r$ is constant on each real gap.
\end{lemma}

\begin{proof}
Off the spectrum, the distance to the compact spectral support is
positive, so the logarithm is bounded on that support. Its integral
against the corresponding probability measure is therefore finite.
If $E_1<E_2$ lie in the same gap, then
$N_r(E_2)-N_r(E_1)=\mu_r((E_1,E_2])=0$, since $\mu_r$ is supported
on $\sigma(H_r)$.
\end{proof}
The \emph{real gap label} of $G_n(r)$ is the value $N_r(E)$ at any
energy $E\in G_n(r)$. This value is independent of the chosen $E$
and equals the density-of-states mass below the gap. This value,
rather than the branch index $n$, is preserved as the gap varies:
\begin{lemma}[Gap label stability; $LE$]
\label{lem:gap-label-constancy}
For $0<r\le a\le1$, the value $N_r(E)$ is locally constant on
the joint resolvent set $E\notin\sigma(H_r)$. It is therefore constant
along open real gap branches and is preserved under limits of cuts
whose limiting cut remains off the spectrum, including at $a=r=1$.
\end{lemma}

\begin{proof}
Near any resolvent cut, norm continuity of $H_r$ gives a common gap.
Continuous functional calculus then makes the lower spectral
projection $P=\mathbf1_{(-\infty,E]}(H_r)$ norm-continuous in the
couplings and independent of the cut within that gap. Bellissard's
invariance of its $K_0$ class under gap-preserving
deformations~\cite[Section~4.2, Theorem~16(iv)]{Bellissard2003}
and the fixed trace give
\begin{align}
 N_r(E)=\tr P=\tr_*[P]
\end{align}
locally constant. Overlapping common-gap neighborhoods give branch
constancy and the asserted preservation at an open limit.
\end{proof}

Now, we show that the transformations in Sec.~\ref{subsec:flatten} preserve the height function:

\begin{lemma}[Height correspondence; $LE$]
\label{lem:height-correspondence}\label{lem:normalization}
Let $0<a,r\le1$ and let $\alpha$ be irrational. For every $t\in\mathbb R$,
\begin{align}\label{eq:height-correspondence}
 h_r\bigl((M/a)\cos t\bigr)&=h_r^W(e^{it})
   =L_{\lambda_1,\lambda_2,\alpha}(e^{it})\in\mathbb R,
\end{align}
where the Lyapunov exponent is normalized per physical lattice step.
\end{lemma}

\begin{proof}
The trace-preserving reduction of Section~\ref{subsec:flatten} gives
\begin{align}
 h_r\bigl(\tfrac{M}{a}\cos t\bigr)
 =\tr_2\!\left(\log\left|
       \tfrac{M}{a}\cos t\,I-\tfrac{M}{a}\operatorname{Re}W
                         \right|\right).
\end{align}
Since $W$ is unitary,
\begin{align}
 2\cos t\,I-(W+W^*)=e^{-it}(e^{it}I-W)(e^{it}I-W^*).
\end{align}
Both factors belong to the commutative algebra generated by $W$.
We first interpret logarithms as extended spectral integrals (or use
continuous logarithmic truncations). Compact spectral support bounds their
positive parts; the Thouless identity below subsequently proves finiteness.
Taking absolute values and logarithms in this sense yields
\begin{align}
 h_r\bigl(\tfrac{M}{a}\cos t\bigr)=\log\frac{M}{2a}+\tr_2(\log|e^{it}I-W|)
                         +\tr_2(\log|e^{it}I-W^*|) =h_r^W(e^{it}).
\end{align}
The second equality uses the reflection $R_1WR_1=W^*$ from
Lemma~\ref{lem:reflection} and invariance of $\tr_2$ under unitary conjugation.

It remains to identify the common height with the Lyapunov exponent.
With $b(x)=\lambda_2\sin(2\pi x)$ and $p(x)=\sqrt{1-b(x)^2}$,
as in Section~\ref{subsec:real-cocycle}, the half-angle formulas give
\begin{align}
 \lambda_1p(x)=\frac{2a}{M}|1+r^2e^{4\pi ix}|.
\end{align}
For $r<1$, the absolutely uniformly convergent series
\begin{align}
\log|\lambda_1p(x)| - \log\left|\frac{2a}{M}\right|= \log|1+r^2e^{4\pi ix}|
 =\operatorname{Re}\sum_{n=1}^{\infty}
       \frac{(-1)^{n+1}r^{2n}e^{4\pi inx}}{n}
\end{align}
has zero integral over $\mathbb T$. At $r=1$ the integrand is
$\log|2\cos(2\pi x)|$, whose singularities have the integrable form
$\log|x-x_0|+O(1)$. Reflection and the double-angle identity give
\begin{align}
 2\int_0^{\pi/2}\log(\sin u)\,du =\int_0^{\pi/2}\log(\sin u\cos u)\,du =\int_0^{\pi/2}\log(\sin u)\,du-\frac\pi2\log2,
\end{align}
so $\log|2\cos(2\pi x)|$ integrates to zero as well, and for every $0<r\le1$,
\begin{align}
 \int_{\mathbb T}\log(\lambda_1p(x))\,dx=\log\frac{2a}{M}.
\end{align}

The CMV representation in \cite[Corollaries~2.13--2.14]{CFO23} (see also \cite{CFLOZ23})
has alternating coefficients $\lambda_1'$ and $b(x)$, with
$\rho$-factors $\lambda_1$ and $p(x)$.
To apply the ergodic CMV formula, use $\mathbb T\times\{0,1\}$
with Haar measure and uniform parity, and the transformation
\begin{align}
 (x,0)&\longmapsto(x,1),\qquad
 (x,1)\longmapsto(x+\alpha,0).
\end{align}
Sampling $\lambda_1'$ on parity $0$ and $b(x+\alpha)$ on parity $1$
gives the stated alternating coefficients. The transformation is
ergodic: its square is irrational rotation on each parity and it
exchanges the two parities. Averaging the two site spectral measures
gives exactly $\nu_r$, the normalized trace measure of the walk.
The Thouless integrability follows from $\lambda_1'<1$,
$\log p\in L^1(\mathbb T)$, and
\begin{align}
 -\log(1-|b(x)|)=-2\log p(x)+\log(1+|b(x)|).
\end{align}
At $r=1$, the preceding logarithmic-cosine calculation gives
$\log p\in L^1(\mathbb T)$; we work on the invariant full-measure
set of phases whose orbits avoid the zeros of $p$.
The CMV Thouless formula~\cite[Theorem~B.1]{FDG}, together with
the correspondence above, therefore gives
\begin{align}
 L_{\lambda_1,\lambda_2,\alpha}(e^{it})
 =2\int\log|e^{it}-w|\,d\nu_r(w)+\log\frac{M}{2a}
 =h_r^W(e^{it}),
\end{align}
where we used the preceding mean formula. The factor two accounts
for the two CMV steps per physical lattice step, with $\nu_r$
normalized as a probability measure.
Finally, $\det A_t(x)=1$ and
$\log^+\|A_t(x)\|\le\log(4/\lambda_1)-\log p(x)$ ensure that
the exponent, and hence the height, is finite.
\end{proof}

\begin{corollary}[Zero spectral height; $LE$]
\label{cor:spectral-baseline}
If $\alpha$ is irrational and $0<r\le a\le1$, then
\begin{align}\label{eq:spectral-baseline}
 h_r(E)&=0\qquad(E\in\sigma(H_r)).
\end{align}
This includes the critical line $r=a$ and the corner $a=r=1$.
\end{corollary}

\begin{proof}
By Cedzich--Fillman--Ong~\cite[Corollary~2.10]{CFO23}, for
$e^{it}\in\sigma(W)$,
\begin{align}
 L_{\lambda_1,\lambda_2,\alpha}(e^{it})
 =\max\left\{0,\log\frac{\lambda_2(1+\lambda_1')} {\lambda_1(1+\lambda_2')}\right\}
 =\max\{0,\log(r/a)\}=0.
\end{align}
The cited result includes $\lambda_1=\lambda_2=1$.
For each $E\in\sigma(H_r)$, the spectral correspondence of
Section~\ref{subsec:flatten} supplies $e^{it}\in\sigma(W)$ with $E=(M/a)\cos t$.
Equation~\eqref{eq:height-correspondence} gives $h_r(E)=0$.
\end{proof}

The trace representation also controls the dependence of the
height on the coupling and energy:
\begin{lemma}[Upper semicontinuity; $L^*$]
\label{lem:height-usc}\label{thm:height-usc}
For fixed $0<a\leq1$ and irrational $\alpha$, the height
$(r,E)\mapsto h_r(E)$ is upper semicontinuous on
$(0,a]\times\mathbb R$.
\end{lemma}

\begin{proof}
We use the regularization argument of \cite[Lemma~3.3]{BorgniaFaust2026}.
Since $H_r$ is norm-continuous on $(0,a]$, continuous functional
calculus shows that, for every $0<\varepsilon\le1$,
\begin{align}
 (r,E)\longmapsto
 \tr\!\left(\log\bigl(|EI-H_r|+\varepsilon I\bigr)\right)
\end{align}
is continuous. As $\varepsilon\downarrow0$, these traces decrease to
$h_r(E)$: on the compact support of $\mu_r$, subtract the logarithmic
kernels from a common upper bound and apply monotone convergence.
Thus
\begin{align}
 h_r(E)=\inf_{0<\varepsilon\le1}
 \tr\!\left(\log\bigl(|EI-H_r|+\varepsilon I\bigr)\right).
\end{align}
An infimum of continuous functions is upper semicontinuous.
\end{proof}

By Lemma~\ref{lem:normalization}, $h_n$ is the maximum Lyapunov
exponent on either lifted gap. To locate it, introduce the resolvent
and its trace for real $E\notin\sigma(H_r)$:
\begin{align}
 \label{eq:resolvent-g}
 R(E,r)&=(E-H_r)^{-1},\qquad
 g(E,r)=\langle\Omega,R(E,r)\Omega\rangle=\tr R(E,r).
\end{align}

\begin{lemma}[The height maximum; $L^*$]\label{lem:gap-maximum}
Every internal gap has a unique maximizing energy $E_n(r)$,
characterized by $g(E_n(r),r)=0$, and $h_n(r)=h_r(E_n(r))>0$.
Its location and height are smooth along an open gap branch.
\end{lemma}

\begin{proof}
Following~\cite[Lemma~3.4]{BorgniaFaust2026}, fix an internal gap
$G_n=(E_-,E_+)$. Corollary~\ref{cor:spectral-baseline} gives
$h_r(E_-)=h_r(E_+)=0$.
To prove continuity from within the gap, split the potential integral
over $(-\infty,E_-]$ and $[E_+,\infty)$. As $E\downarrow E_-$,
the first integrand decreases to $\log|E_--x|$; subtracting from a
common upper bound permits monotone convergence to the finite
endpoint integral. The second integrand stays away from its singularity,
so dominated convergence applies. The other endpoint is analogous.

For $E\in G_n$, distance to the spectrum is positive, and
differentiation under the integral gives
\begin{align}
 \partial_{E}h_r(E)&=\tr(E-H_r)^{-1}=g(E,r),\qquad
 \partial_{E}^{2}h_r(E)=-\tr(E-H_r)^{-2}<0.
\end{align}
Strict concavity and the zero endpoint values give a unique positive
interior maximum, characterized by $\partial_{E}h_r(E_n(r))=0$.
Since the resolvent is smooth and $\partial_{E} g<0$, the implicit
function theorem makes $E_n$ and hence $h_n$ smooth along the branch.
\end{proof}

At the gap maximum, the chain rule reduces the radial derivative
to a resolvent pairing.
\begin{lemma}[$L^*$]\label{lem:derivative-bound}
At $E=E_n(r)$, let $R=R(E,r)$. Then
\begin{align}
 \label{eq:height-derivative}
 r\frac{d}{dr}h_n(r)&=-\langle\Omega,R\,r(\partial_{r} H_r)\Omega\rangle.
\end{align}
\end{lemma}

\begin{proof}
The spectral gap bounds $(E-H_r)^2$ below locally. Differentiating
the logarithm under the trace and using cyclicity therefore gives
\begin{align}
 \partial_{r} h_r(E)
 &=\tfrac12\tr\!\left((E-H_r)^{-2}\partial_{r}(E-H_r)^2\right)
 =-\tr(R\,\partial_{r} H_r).
\end{align}
At the maximum, the chain rule, $g(E,r)=0$, and $RH_r=ER-I$ yield
\begin{align}
 \label{eq:maximum-resolvent}
 \frac{d}{dr}h_n(r)&=\partial_{r}h_r(E_n(r)),\qquad
 \langle\Omega,RH_r\Omega\rangle=Eg(E,r)-1=-1.
\end{align}
\end{proof}

Now, our goal is to show that the derivative in \eqref{eq:height-derivative}
is nonnegative. The maximum height of an open gap then remains positive,
so that the spectral vanishing in Corollary~\ref{cor:spectral-baseline} prevents
the gap from closing. 

\begin{theorem}[Height monotonicity; $L^*$]\label{thm:height-monotonicity}
Fix an irrational $\alpha$ and $0<a\leq1$. Along every open internal-gap branch of $H_r$ with
$0<r<a$,
\begin{align}
 \label{eq:monotonicity-bound}
 0&\leq r\frac{d}{dr}h_n(r)\leq2.
\end{align}
\end{theorem}

\begin{remark}[Trajectory to criticality]
The derivative is taken with $a$ fixed. Each choice of $a$ selects
the trajectory $(a,r)$ with $r\uparrow a$; its endpoint $r=a$ is
the critical point, where $\lambda_1=\lambda_2$.
\end{remark}

\subsection{Proof of monotonicity theorem}\label{subsec:cone-estimate}
Our proof of Theorem~\ref{thm:height-monotonicity}
follows a similar approach to the one found in~\cite{BorgniaFaust2026}: 
\begin{enumerate}
    \item Express the radial derivative at the source through the bounded sector parity operator.
    \item Use stationarity to prove that the compressed sector parity and symmetrized Hamiltonian commute on the centered cyclic space.
    \item Use their common resolvent eigenline to bound the pairing in Lemma~\ref{lem:derivative-bound}.
\end{enumerate}

\subsubsection{Sector Parity}
The derivative in Lemma~\ref{lem:derivative-bound} can be expressed
geometrically through a sector parity operator.

\begin{definition}[Canonical magnetic basis]\label{def:magnetic-basis}
For $j,k\in\mathbb Z$, denote the canonical basis in the fixed magnetic coordinates by
\begin{align}
 \label{eq:magnetic-basis}
 e_{j,k}&=e^{\pi i\alpha k(j+1)}U^jV^k\Omega,
 \qquad \Omega=e_{0,0}.
\end{align}
The canonical trace makes this basis orthonormal; it is complete
because $U,V$ generate $\mathcal A$. Axes and quadrants below refer
to the index lattice $(j,k)$. In particular, for the sign index
$\varepsilon\in\{-1,1\}$, the four axis vectors adjacent to $\Omega$ are
\begin{align}
 e_{0,\varepsilon}&=e^{\varepsilon\pi i\alpha}V^{\varepsilon}\Omega,
 \qquad e_{\varepsilon,0}=U^{\varepsilon}\Omega.
\end{align}
\end{definition}

\begin{definition}[Sector parity]\label{def:harper}\label{def:sector-parity}
On the canonical magnetic basis, define
\begin{align}
 \label{eq:cone-observable}
 L_z e_{j,k}&=-\operatorname{sgn}(jk)e_{j,k},
 \qquad \operatorname{sgn}(0)=0.
\end{align}
Thus $L_z$ has opposite signs in adjacent quadrants and vanishes on
the axes. It is independent of $r$, with $L_z=L_z^*$,
$\|L_z\|\leq1$, and $L_z\Omega=0$.
\end{definition}

The connection with the radial derivative comes from the phase twists
$U\mapsto e^{ik_x}U$, $V\mapsto e^{ik_y}V$.
\begin{lemma}[$L^*$]\label{lem:radial-cone}\label{lem:twist-derivative}
With the twist derivatives evaluated at $k_x=k_y=0$,
\begin{align}
 r\partial_{r}H_r&=(I-\partial_{k_x}\partial_{k_y})H_r,
 \qquad
 r(\partial_{r}H_r)\Omega=(I-L_z)H_r\Omega.
 \label{eq:radial-cone}
\end{align}
\end{lemma}

\begin{proof}
In the magnetic basis, the six terms of \eqref{eq:harper} give
\begin{align}
 H_r\Omega&=\sum_{\varepsilon=\pm1}
 \Bigl[e_{\varepsilon,-\varepsilon}-r^2e_{\varepsilon,\varepsilon}
 +r\bigl(a^{-1}e^{-\varepsilon\pi i\alpha}-ae^{\varepsilon\pi i\alpha}\bigr)e_{0,\varepsilon}\Bigr].
 \label{eq:source-hops}
\end{align}
The twists multiply a $(j,k)$ monomial by $e^{i(jk_x+kk_y)}$,
so $\partial_{k_x}\partial_{k_y}$ has multiplier $-jk$.
On the three direction pairs $(1,-1)$, $(0,1)$, and $(1,1)$,
$I-\partial_{k_x}\partial_{k_y}$ therefore has weights $0,1,2$,
matching the radial powers $1,r,r^2$. This proves the first identity.
At these six endpoints $jk\in\{-1,0,1\}$, hence
$jk=\operatorname{sgn}(jk)$ and the mixed derivative agrees with
$L_z$ on $H_r\Omega$.
\end{proof}

At $E=E_n(r)$, write $R=R(E,r)$. Lemma~\ref{lem:derivative-bound}
and \eqref{eq:maximum-resolvent} now give
\begin{align}
 \label{eq:derivative-target}
 r\frac{d}{dr}h_n(r)
 &=-\langle\Omega,R(I-L_z)H_r\Omega\rangle
 =1+\langle\Omega,RL_zH_r\Omega\rangle.
\end{align}
To place this derivative in $[0,2]$, it suffices to prove
\begin{align}
 \label{eq:target-bound}
 \bigl|\langle\Omega,RL_zH_r\Omega\rangle\bigr|&\leq1
 \qquad\text{when }g(E,r)=0.
\end{align}

\subsubsection{Stationarity and parity conservation}
The stationarity argument of~\cite[Lemma~2.2]{BorgniaFaust2026}
uses two commuting Hamiltonian actions that agree at the source.
We realize them by left and right multiplication in the trace space.

\begin{definition}[Cyclic source space and symmetric trace actions]
An algebra element $F\in\mathcal A$ represents the vector $F\Omega$
in $\mathscr H$. Recall the trace inner product from Section~\ref{subsec:flatten}:
$\langle F\Omega,G\Omega\rangle=\tr(F^*G)$ for $F,G\in\mathcal A$.
Left multiplication sends $F\Omega$ to $(H_rF)\Omega$;
right multiplication sends it to $(FH_r)\Omega$. On this dense set
of vectors, define their half-sum and half-difference by
\begin{align}
 H_{r,+}(F\Omega)&:=\tfrac12(H_rF+FH_r)\Omega,\qquad
 H_{r,-}(F\Omega):=\tfrac12(H_rF-FH_r)\Omega.
\end{align}
As before, $H_r$ by itself denotes the left action. The cyclic
source space is the closure, in the trace Hilbert space, of the
polynomial source vectors:
\begin{align}
 \mathcal V_r&:=\overline{\{f(H_r)\Omega:f\text{ a complex polynomial}\}}
 \subset\mathscr H.
\end{align}
\end{definition}

\begin{lemma}[Stationarity; $L^*$]\label{lem:walk-stationarity}\label{lem:scalar-stationarity}
The operators $H_{r,\pm}$ extend boundedly and self-adjointly to
$\mathscr H$. The space $\mathcal V_r$ is invariant under $H_r$ and
$H_{r,+}$, and
\begin{align}
 \label{eq:cyclic-stationarity}
 H_{r,-}|_{\mathcal V_r}&=0,\qquad
 H_{r,+}|_{\mathcal V_r}=H_r|_{\mathcal V_r},\qquad
 H_{r,+}\Omega=H_r\Omega.
\end{align}
\end{lemma}

\begin{proof}
Left and right multiplication by $H_r$ are bounded self-adjoint
actions. They commute by associativity and agree on $\Omega=I$.
Thus $H_{r,\pm}$ commute with $H_r$, and for every polynomial $f$,
\begin{align}
 H_{r,-}f(H_r)\Omega
 &=f(H_r)H_{r,-}\Omega=0,\\
 H_{r,+}f(H_r)\Omega
 &=f(H_r)H_{r,+}\Omega=H_rf(H_r)\Omega\in\mathcal V_r.
\end{align}
Since $H_rf(H_r)\Omega$ is again a polynomial source vector,
boundedness and polynomial density extend both the identities and
invariance to $\mathcal V_r$.
\end{proof}

\begin{remark}[Relation to the magnetic inverse; $W$]\label{remark:maginv}
Stationarity requires only two commuting actions that agree on
$\Omega$; a conjugation formula is not needed for the proof.
In this case, up to the gauge transformation, the right action does correspond to $\bar{K}_r$ of Lemma~\ref{lem:block-symmetry}, after
normalization and an antiunitary change of representation.
Indeed, $F\Omega\mapsto F^*\Omega$ is antiunitary and exchanges
left and right multiplication by $H_r$; combine this map with
$G_r\mathcal C$, which conjugates $(M/a)\bar K_r$ to $H_r$.
\end{remark}

It is on the centered cyclic space consisting of source vectors orthogonal to
$\Omega$ that we can establish the desired conservation law. Let $\Pi_0$ be its orthogonal projection in $\mathscr H$:
\begin{align}
 \mathcal V_r^0&:=\mathcal V_r\cap\Omega^\perp,\qquad
 \Pi_0v=v-\langle\Omega,v\rangle\Omega\quad(v\in\mathcal V_r).
\end{align}
The formula for $\Pi_0v$ is stated on $\mathcal V_r$; all compressed
operators below act on $\mathcal V_r^0$.

\begin{lemma}[Sector parity conservation; $L^*$]\label{lem:axis-terms}\label{lem:cone}
On $\mathcal V_r^0$,
\begin{align}
 \label{eq:cone-identities}
 [\Pi_0L_z\Pi_0,\Pi_0H_{r,+}\Pi_0]&=0.
\end{align}
\end{lemma}

\begin{proof}
We follow the boundary cancellation in
\cite[Lemma~2.4, (2.19)--(2.21)]{BorgniaFaust2026}.
Here $H_r$ has additional hopping directions; their contribution
is the finite corner correction $F_{\mathrm c}$ defined below.
We use the left and right actions directly, as in
Remark~\ref{remark:maginv}.

\emph{1. Boundary balance}
Express each boundary contribution through $H_{r,-}$, then use
stationarity to cancel it. Their cone boundaries are diagonals;
here they are the coordinate axes, so we compute the coefficients
in our fixed basis.

Write $(s,t)\in\mathbb Z^2$ for a hopping increment: its source is
$e_{j,k}$ and its target is $e_{j+s,k+t}$. The coefficient
$\langle e_{s,t},H_r\Omega\rangle$ is given by
\eqref{eq:source-hops}; it vanishes except at
$(\varepsilon,-\varepsilon)$, $(\varepsilon,\varepsilon)$, and
$(0,\varepsilon)$, with $\varepsilon=\pm1$.
The relation $UV=e^{-2\pi i\alpha}VU$ and the phases in
\eqref{eq:magnetic-basis} give left and right multiplication factors
$e^{\mp\pi i\alpha(sk-tj)}$. Taking their half-sum and half-difference as in the definition of $H_{r,\pm}$ gives
\begin{align}
 \langle e_{j+s,k+t},H_{r,+}e_{j,k}\rangle
 &=\cos\bigl(\pi\alpha(sk-tj)\bigr)\langle e_{s,t},H_r\Omega\rangle,\\
 \langle e_{j+s,k+t},H_{r,-}e_{j,k}\rangle
 &=-i\sin\bigl(\pi\alpha(sk-tj)\bigr)\langle e_{s,t},H_r\Omega\rangle.
 \label{eq:symmetric-hopping-coefficients}
\end{align}

For a source $e_{0,k}$ with $k\ne0$, a hop to an off-axis target
has $s=\pm1$. Its $H_{r,+}$ and $H_{r,-}$ factors are respectively
$\cos(\pi\alpha k)$ and $-is\sin(\pi\alpha k)$, and its target
parity is $-s\operatorname{sgn}(k)$. For a source $e_{j,0}$ with
$j\ne0$, the corresponding factors are $\cos(\pi\alpha j)$ and
$it\sin(\pi\alpha j)$, and the target parity is
$-t\operatorname{sgn}(j)$. Thus the two scalar conversions are
\begin{align}
 -s\operatorname{sgn}(k)\cos(\pi\alpha k)
 &=\bigl[-i\operatorname{sgn}(k)\cot(\pi\alpha k)\bigr]
   \bigl[-is\sin(\pi\alpha k)\bigr],\\
 -t\operatorname{sgn}(j)\cos(\pi\alpha j)
 &=\bigl[i\operatorname{sgn}(j)\cot(\pi\alpha j)\bigr]
   \bigl[it\sin(\pi\alpha j)\bigr].
\end{align}
Multiplying by the common hopping coefficient yields boundary balance in \cite{BorgniaFaust2026} for off-axis targets.

\emph{2. Corner correction for $H_r$.}
When a diagonal hop lands on the other axis, its target parity is
zero, requiring a correction on the four
nearest axis sites,
\begin{align}
 F_{\mathrm c}e_{j,k}&:=\cos(\pi\alpha)\,\delta_{|j|+|k|,1}
 \bigl(e_{k,j}+r^2e_{-k,-j}\bigr).
 \label{eq:corner-actions}
\end{align}
This defines a finite-rank operator with symmetric reverse hopping terms, such that $F_{\mathrm c}=F_{\mathrm c}^*$ and
$F_{\mathrm c}\Omega=0$.
%
Consequently the full identities are
\begin{align}
 L_zH_{r,+}e_{0,k}
 &=-i\operatorname{sgn}(k)\cot(\pi\alpha k)H_{r,-}e_{0,k}
   +F_{\mathrm c}e_{0,k}\quad(k\ne0),\label{eq:vertical-axis}\\
 L_zH_{r,+}e_{j,0}
 &=i\operatorname{sgn}(j)\cot(\pi\alpha j)H_{r,-}e_{j,0}
   +F_{\mathrm c}e_{j,0}\quad(j\ne0).\label{eq:horizontal-axis}
\end{align}
And, $F_{\mathrm c}$ is the additional term in our boundary identities.

\emph{3. Stationary cancellation.}
Let $\Pi_{\mathrm{ax}}$ be the orthogonal projection onto the closed
span of $\{e_{j,k}:jk=0\}$. Each hop changes either index by at most
one. Therefore hops with both endpoints off the axes have equal
parity at their endpoints, while hops between two axis sites have
parity zero at both. Only hops with exactly one axis endpoint remain:
\begin{align}
 [L_z,H_{r,+}]
 &=L_zH_{r,+}\Pi_{\mathrm{ax}}-\Pi_{\mathrm{ax}}H_{r,+}L_z.
 \label{eq:axis-commutator}
\end{align}

Take $v,w\in\mathcal V_r^0$, each obtained by applying a polynomial
in $H_r$ to $\Omega$. They have finite support in the basis $e_{j,k}$ and satisfy
$H_{r,-}v=H_{r,-}w=0$. For every axis basis vector,
\begin{align}
 \langle v,H_{r,-}e_{j,k}\rangle
 &=\langle H_{r,-}v,e_{j,k}\rangle=0.
\end{align}
Hence each cotangent term in \eqref{eq:vertical-axis} and
\eqref{eq:horizontal-axis} vanishes before summation. Irrationality
makes each cotangent finite; no bound uniform in $j,k$ is needed.
The origin column is absent because $w\perp\Omega$, and so
\begin{align}
 \langle v,L_zH_{r,+}\Pi_{\mathrm{ax}}w\rangle
 &=\sum_{\substack{jk=0\\(j,k)\ne(0,0)}}
   \langle v,L_zH_{r,+}e_{j,k}\rangle\langle e_{j,k},w\rangle
 =\langle v,F_{\mathrm c}w\rangle.
\end{align}
Subtracting the adjoint pairing in \eqref{eq:axis-commutator} now gives
\begin{align}
 \label{eq:stationary-commutator}
 \langle v,[L_z,H_{r,+}]w\rangle
 &=\langle v,F_{\mathrm c}w\rangle-\langle F_{\mathrm c}v,w\rangle=0.
\end{align}
\emph{4. Centered compression.}
For $v,w\in\mathcal V_r^0$, invariance gives
$H_{r,+}v,H_{r,+}w\in\mathcal V_r$. Their $\Pi_0$-projections
remove only their $\Omega$-components, which pair to zero with
$L_zv,L_zw$ because $L_z\Omega=0$. Explicitly,
\begin{align}
 \langle v,\Pi_0L_z\Pi_0H_{r,+}\Pi_0w\rangle
 &=\langle L_zv,\Pi_0H_{r,+}w\rangle
 =\langle L_zv,H_{r,+}w\rangle,\\
 \langle v,\Pi_0H_{r,+}\Pi_0L_z\Pi_0w\rangle
 &=\langle\Pi_0H_{r,+}v,L_zw\rangle
 =\langle H_{r,+}v,L_zw\rangle.
\end{align}
Their difference is the pairing in \eqref{eq:stationary-commutator}.
Centered polynomial vectors are dense in $\mathcal V_r^0$, and
all operators in these pairings are bounded. And, the compressed
commutator vanishes on $\mathcal V_r^0$.
\end{proof}

\subsubsection{Sector Parity Eigenvalues}
The condition $g(E,r)=0$ means that the resolvent vector is
orthogonal to the source. The following computation shows exactly
why removing the source then creates a simple eigenvalue.

\begin{lemma}[ $L^*$]\label{lem:gap-resolvent}
For real $E\notin\sigma(H_r)$, the compression $\Pi_0H_{r,+}\Pi_0$
on $\mathcal V_r^0$ has eigenvalue $E$ if and only if $g(E,r)=0$.
In that case its $E$-eigenspace is $\mathbb C R(E,r)\Omega$.
\end{lemma}

\begin{proof}
Write $R=R(E,r)$. Functional calculus gives $R\Omega\in\mathcal V_r$,
so stationarity and the resolvent identity give
\begin{align}
 (E-H_{r,+})R\Omega&=(E-H_r)R\Omega=\Omega,\qquad
 \langle\Omega,R\Omega\rangle=g(E,r).
\end{align}
If $g(E,r)=0$, then $R\Omega$ is centered, and
\begin{align}
 \Pi_0R\Omega&=R\Omega,\qquad
 \Pi_0H_{r,+}\Pi_0R\Omega
 =\Pi_0(ER\Omega-\Omega)=ER\Omega.
 \label{eq:projected-resolvent}
\end{align}
Thus $R\Omega\ne0$ is an $E$-eigenvector of the compression.

Conversely, let $0\ne v\in\mathcal V_r^0$ be a compressed
$E$-eigenvector. Since $(E-H_{r,+})v\in\mathcal V_r$, its vanishing
$\Pi_0$-projection means that it is a multiple of $\Omega$.
Stationarity and inversion of $E-H_r$ give, for some scalar
$\gamma\in\mathbb C$,
\begin{align}
 \Pi_0H_{r,+}\Pi_0v=Ev
 &\quad\Longleftrightarrow\quad (E-H_r)v=\gamma\Omega
 \quad\Longleftrightarrow\quad v=\gamma R\Omega.
\end{align}
Here $\gamma\ne0$ because $v\ne0$. Centering now gives
\begin{align}
 0&=\langle\Omega,v\rangle
 =\gamma\langle\Omega,R\Omega\rangle=\gamma g(E,r),
\end{align}
so $g(E,r)=0$. Every eigenvector is a multiple of $R\Omega$,
which also proves that the eigenspace is one-dimensional.
\end{proof}

\begin{corollary}[$L^*$]\label{cor:resolvent-parity}
At a real resolvent energy with $g(E,r)=0$, write $R=R(E,r)$.
There is an $\ell\in[-1,1]$ such that
\begin{align}
 \label{eq:parity-pairing-bound}
 (\Pi_0L_z\Pi_0)R\Omega&=\ell R\Omega,\qquad
 \bigl|\langle\Omega,RL_zH_{r,+}\Omega\rangle\bigr|=|\ell|\leq1.
\end{align}
\end{corollary}

\begin{proof}
By Lemmas~\ref{lem:cone} and~\ref{lem:gap-resolvent}, the self-adjoint
contraction $\Pi_0L_z\Pi_0$ preserves the line $\mathbb C R\Omega$,
giving the first assertion. And, $\tr H_r=0$ implies
$H_{r,+}\Omega\in\mathcal V_r^0$. Using $R=R^*$ and $g(E,r)=0$ gives
\begin{align}
 \langle\Omega,RL_zH_{r,+}\Omega\rangle
 &=\langle\Pi_0L_z\Pi_0R\Omega,H_{r,+}\Omega\rangle
 =\ell\bigl(Eg(E,r)-1\bigr)=-\ell.
\end{align}
Then, $\vert\vert L_{z}\vert\vert\leq 1$ gives $\vert \ell\vert \leq1$
\end{proof}

\begin{proof}[Proof of Theorem~\ref{thm:height-monotonicity}]
At $E=E_n(r)$, Lemma~\ref{lem:gap-maximum} gives $g(E,r)=0$.
Since $H_{r,+}\Omega=H_r\Omega$, Corollary~\ref{cor:resolvent-parity}
proves \eqref{eq:target-bound} with $|\ell|\leq1$.
Substituting into
\eqref{eq:derivative-target} gives
\begin{align}
 \label{eq:height-identity}
 r\frac{d}{dr}h_n(r)
 &=-\langle\Omega,R(I-L_z)H_r\Omega\rangle=1-\ell\in[0,2].
\end{align}
And, the maximum height is a non-decreasing function in $r$.
\end{proof}

\section{Proof of Theorem~\ref{thm:classification}}
\label{sec:fullproof}\label{subsec:gap-continuation}
Fixing the irrational frequency $\alpha$, coupling-exchange
duality reduces the proof to $0<r\leq a\leq1$. Section~\ref{subsec:flatten}
identifies internal gaps of $H_r$ with conjugate gaps of $W$. Thus, Proposition~\ref{prop:gap-persistence} reduces their opening to
continuation from known gaps at fixed $a$.

The points $\cos t=\pm1$ require separate treatment: gaps through
$1,-1$ correspond to exterior real components, to which internal-gap
persistence does not apply. For $a<1$, Theorem~\ref{thm:interior}
already opens all gaps for $r<a$, but the lean verification is simpler if we continue the seed gaps from Appendix~\ref{app:interior-seeds} to
$r=a$ and check the gaps through $1,-1$ directly. On $a=1$,
Section~\ref{Sec:two} supplies only odd-winding seeds and excludes
even windings for $r<1$. Here $1,-1$ are included in the spectrum. So, we
continue the odd gaps and retain the even-gap obstruction at the corner.

\smallskip\noindent\emph{Gap labels.}
Recall the increasing rotation-number lift normalized by
\begin{align}
 \label{eq:dos-rotation}
 \rho(t)&=\nu_r\bigl(\{e^{iu}:0<u\leq t\}\bigr),
 \qquad 0\leq t\leq2\pi.
\end{align}
Equation~\eqref{eq:dos-rotation} is the DOS normalization of
\cite[proof of Theorem~4.3]{CL26} ($LE$) for $a,r<1$ and extends
to $a=1$, $r<1$ by continuity of the rotation number and the
integrated DOS. The required atomless DOS follows from
the trace reduction in Section~\ref{subsec:flatten} and
Corollary~\ref{cor:spectral-baseline}. At $a=r=1$,
\eqref{eq:dos-rotation} defines the gap values.

For $E=(M/a)\cos t$ and $0<t<\pi$,  Lemma~\ref{lem:reflection} gives
\begin{align}
 \label{eq:real-unitary-labels}
 \rho(t)&=\frac{1-N_r(E)}2,\qquad
 \rho(2\pi-t)=\frac{1+N_r(E)}2.
\end{align}
By Lemma~\ref{lem:cosine} and \eqref{eq:harper-spectrum}, an
internal real gap with label $s$ therefore lifts to gaps with
values $(1-s)/2$ and $(1+s)/2$ on the respective semicircles.
Conversely, $\rho_*\in(0,1)\setminus\{1/2\}$ gives the real label
$s=|1-2\rho_*|$.
Lemma~\ref{lem:gap-label-constancy} preserves $s$, and hence each
unitary value on its semicircle, along open branches and at open
limits. To realize both values of a fixed winding in \eqref{eq:labels},
we apply continuation to each $\rho_*$ separately.

\begin{proposition}[Gap persistence; $L^*$]
\label{prop:gap-persistence}
Fix $0<a\leq1$. An internal gap open at $r_0\in(0,a)$ extends to
an open branch on $[r_0,a]$ with the same real
label. Along this branch,
\begin{align}
 h_n(r)&\geq h_n(r_0)>0.
\end{align}
\end{proposition}
\begin{proof}
We use the compactness argument of
\cite[proof of Theorem~3.10]{BorgniaFaust2026}.
Fix the initial label $s\in(0,1)$ and target $r_1\in(r_0,a]$.
Consider all resolvent pairs $(r,E)$, $r\in[r_0,r_1]$, satisfying
\begin{align}
 g(E,r)&=0,\qquad N_r(E)=s,\qquad h_r(E)\geq h_n(r_0)>0.
\end{align}
Lemma~\ref{lem:gap-maximum} supplies the initial pair.
Since $0<s<1$, their energies satisfy $|E|\leq\|H_r\|$ and are
uniformly bounded. For any convergent sequence of these pairs,
Lemma~\ref{lem:height-usc} gives
\begin{align}
 \label{eq:persistence-positive-limit}
 h_{r_*}(E_*)&\geq\limsup_{j\to\infty}h_{r_j}(E_j)
 \geq h_n(r_0)>0.
\end{align}
Thus $E_*$ remains off the spectrum by
Corollary~\ref{cor:spectral-baseline}. Resolvent continuity and
Lemma~\ref{lem:gap-label-constancy} preserve $g=0$ and $N_r(E)=s$
at the limit, so the set of pairs is compact.
Its largest coupling must be $r_1$: otherwise
Lemma~\ref{lem:gap-maximum} and Theorem~\ref{thm:height-monotonicity}
would extend a maximizing pair to larger $r$ with the same label
and height bound.

These gaps form the original branch. A fixed cut in any selected gap
stays in the resolvent with label $s$ for nearby $r$, by
Lemma~\ref{lem:gap-label-constancy}. Faithfulness gives
$\operatorname{supp}\mu_r=\sigma(H_r)$, so distinct gaps have
distinct IDS values; the cut therefore belongs to the selected
gap at every nearby coupling. Equation~\eqref{eq:real-unitary-labels}
preserves both unitary values.

This includes $r_1=a=1$: norm continuity and zero spectral height
hold at the corner, while monotonicity is used only for $r<a$.
\end{proof}

\subsection{Proof of Theorem~\ref{thm:classification}\textup{(i)}}
\label{subsec:proof-interior}
\begin{proof}
Fix $0<a<1$. Theorem~\ref{thm:interior} settles $r<a$; only
$r=a$ remains.

\smallskip\noindent\emph{Zero winding.}
At $r=a<1$, the matrices in \eqref{eq:real-cocycle} are diagonal
for $t=0,\pi$, with positive and negative entries, respectively.
By \eqref{eq:half-angle},
\begin{align}
 \label{eq:diagonal-log-mean}
 \int_{\mathbb T}\log|(A_t(x))_{11}|\,dx
 &=\log a+\frac12\int_{\mathbb T}
 \log\frac{1\mp b(x)}{1\pm b(x)}\,dx
 =\log a<0,\qquad t=0,\pi.
\end{align}
The signs correspond to $t=0,\pi$, respectively; the ratio integral
vanishes because $b(x+\tfrac12)=-b(x)$.
Continuity of the logarithms and unique ergodicity give uniform
rates $\log a,-\log a$ for the reciprocal diagonal moduli.
Thus the cocycles are uniformly hyperbolic, and the spectral
criterion of Section~\ref{Sec:one} opens the gaps through $1,-1$
with values $0,1/2$.

\smallskip\noindent\emph{Nonzero winding.}
For each $m\ne0$ and either representative $\rho_*\in(0,1)$
satisfying \eqref{eq:labels},
Theorem~\ref{thm:interior} supplies a seed at $0<r_0<a$
(alternatively, use Proposition~\ref{prop:appendix-interior-seeds}).
Its real label $|1-2\rho_*|\in(0,1)$ makes the scalar gap internal.
Proposition~\ref{prop:gap-persistence} continues it to $r=a$,
preserving $\rho_*$ by \eqref{eq:real-unitary-labels}.
\end{proof}

\subsection{Proof of Theorem~\ref{thm:classification}\textup{(ii)}}
\label{subsec:proof-boundary}
\begin{proof}
Take $a=1$. Proposition~\ref{prop:boundary-even-closure} excludes
even windings for $r<1$. In particular, $1,-1$ are spectral,
since a gap through either point would have winding zero.
Norm continuity of $W$ retains both points at $r=1$.

\smallskip\noindent\emph{Odd windings.}
Fix a target $0<r\leq1$. For each odd $m$ and either permitted
$\rho_*$, Proposition~\ref{prop:odd-gap-seeds} and
\eqref{eq:half-angle} supply a seed at $0<r_0<r$.
Its scalar gap is internal by \eqref{eq:real-unitary-labels},
so Proposition~\ref{prop:gap-persistence} continues it to the target
with the same $\rho_*$, including at $r=1$.

\smallskip\noindent\emph{No even windings at the corner.}
Choose $z=e^{it}$ in a gap at $r=1$; since $z\ne\pm1$, it lies
in one open semicircle. Norm continuity keeps $z$ in the resolvent
for $r<1$ close to $1$. Then \eqref{eq:harper-spectrum} and
Lemma~\ref{lem:gap-label-constancy} give
\begin{align}
 E_r&=2(1+r^2)\operatorname{Re}z,\qquad
 N_r(E_r)=N_1(4\operatorname{Re}z).
\end{align}
Equation~\eqref{eq:real-unitary-labels} therefore preserves
$\rho(t)$ on this semicircle. The corner gap has exactly the label
of the nearby edge gap, whose winding is odd by
Proposition~\ref{prop:boundary-even-closure} and unique by
\eqref{eq:labels}.
\end{proof}

\section*{Acknowledgements}
L. Li thanks Xianzhe Li for  the statement of \cite[Lemma 8.6]{LiXuZhou2026} and explanation of the related literature.

Work by D. S. B. was supported by the Low-Energy AI-Informed Phase Transitions (UC-LEAP) project as part of the University of California AI Science at Scale initiative.
Part of this work was conducted while M. F. was a research fellow at the Simons Institute for the Theory of Computing.

\ \\
\textbf{AI Acknowledgment}
The authors acknowledge the use of ChatGPT to accelerate the completion of this project. The idea of adapting the height function argument developed in \cite{BorgniaFaust2026} to the UAMO was initiated by L. Li; he suggested that we work with AI to search for an isospectral element of the magnetic operator algebra, Eq.~\eqref{eq:harper}. We provided ChatGPT with L. Li's notes on the subcritical dry ten martini problem for the UAMO in order to generate the arguments used in the proof of Theorem~\ref{thm:interior}. The observation in Proposition~\ref{prop:boundary-even-closure} about the closure of even gaps was made in discussion with ChatGPT. AI was then used to generate the perturbative argument in Section~\ref{subsec:boundary} showing that the odd seed gaps are indeed open. The adaptation of the height function from~\cite{BorgniaFaust2026} was then completed in discussion with ChatGPT. 

Much of the initial drafting was done by ChatGPT and then rewritten, edited, and organized by the authors into the manuscript. The authors take full responsibility for the validity of all mathematical arguments and conclusions in the manuscript. We also acknowledge the use of \href{https://relace.ai}{Relace}'s \href{https://jacquard.dev}{Jacq} platform and ChatGPT for assisting us in the formalization of the proof in Lean and generating Appendix~\ref{app:lean-audit}.

\appendix
\section{Physical profiles and Fourier duality}\label{app:duality}

For a transfer profile $v=(v_1,v_2)^T$, the associated same-site
profile is
\begin{align}
 (\mathcal R_\varkappa v)(x)
 =\begin{pmatrix}v_1(x)\\\varkappa v_2(x+\alpha)\end{pmatrix}.
\end{align}
This map is unitary on $L^2(\T;\C^2)$: translation preserves Lebesgue
measure and $|\varkappa|=1$. In particular, for every fixed
$v\in\C^2$, the columnwise conversion of $Z$ to $\Phi$ satisfies
\begin{align}
 \int_{\T}|\Phi(x)v|^2\,dx=\int_{\T}|Z(x)v|^2\,dx.
\end{align}
This is an integrated identity; it does not assert pointwise
invertibility of the same-site matrix $\Phi(x)$.

We give an explicit verification of the Fourier duality identity.
Write $\varkappa=e^{2\pi i\xi}$ and set
\begin{align}
K_F=\frac{1}{\sqrt{2}}
\begin{pmatrix}
1&i\\
-i&-1
\end{pmatrix},
\qquad
(\mathcal Ff)_n
=\int_{\mathbb T}e^{-2\pi inx}f(x)\,dx.
\end{align}
Here $\mathcal F:L^2(\mathbb T;\mathbb C^2)\to
\ell^2(\mathbb Z;\mathbb C^2)$ acts componentwise, and
$K_F^{-1}=K_F^*=K_F$.

For a continuous same-site profile $f$, the associated physical sequence is
$u_n=\varkappa^n f(\theta+n\alpha)$. Thus the profile shift is
\begin{align}
(\mathscr S_{\lambda_1,\varkappa}f)(x)
=
\begin{pmatrix}
\lambda_1\varkappa^{-1}f_+(x-\alpha)-\lambda_1'f_-(x)\\
\lambda_1\varkappa f_-(x+\alpha)+\lambda_1'f_+(x)
\end{pmatrix},
\end{align}
and
\begin{align}
\mathscr W_\varkappa
=\mathscr S_{\lambda_1,\varkappa}Q_{\lambda_2},
\end{align}
where $Q_{\lambda_2}$ denotes multiplication by
$Q_{\lambda_2}(x)$.

Set $V=K_F\mathcal F$. We first compute the image of the
profile shift. Since
\begin{align}
\mathcal F[f(\,\cdot\pm\alpha)]_n
=e^{\pm2\pi in\alpha}\widehat f_n,
\end{align}
the operator
$\mathcal F\mathscr S_{\lambda_1,\varkappa}\mathcal F^{-1}$
acts at site $n$ by
\begin{align}
D_n=
\begin{pmatrix}
\lambda_1e^{-2\pi i(\xi+n\alpha)}&-\lambda_1'\\
\lambda_1'&\lambda_1e^{2\pi i(\xi+n\alpha)}
\end{pmatrix}.
\end{align}
Direct multiplication gives
\begin{align}
K_F
\begin{pmatrix}
\lambda e^{it}&-\lambda'\\
\lambda'&\lambda e^{-it}
\end{pmatrix}
K_F
=
\begin{pmatrix}
\lambda\cos t+i\lambda'&-\lambda\sin t\\
\lambda\sin t&\lambda\cos t-i\lambda'
\end{pmatrix}
=
Q_\lambda\!\left(\frac{t}{2\pi}\right).
\end{align}
Taking $t=-2\pi(\xi+n\alpha)$ therefore yields
\begin{align}
V\mathscr S_{\lambda_1,\varkappa}V^{-1}
=Q_{\lambda_1,-\alpha,-\xi},
\end{align}
where
\begin{align}
(Q_{\lambda_1,-\alpha,-\xi}a)_n
=Q_{\lambda_1}(-\xi-n\alpha)a_n.
\end{align}

Next, the same matrix identity gives
\begin{align}
K_FQ_{\lambda_2}(x)K_F
=
\begin{pmatrix}
\lambda_2e^{2\pi ix}&-\lambda_2'\\
\lambda_2'&\lambda_2e^{-2\pi ix}
\end{pmatrix}.
\end{align}
Using
\begin{align}
\mathcal F[e^{2\pi ix}f]_n=\widehat f_{n-1},
\qquad
\mathcal F[e^{-2\pi ix}f]_n=\widehat f_{n+1},
\end{align}
and the fact that $K_F$ commutes with the componentwise
Fourier transform, we obtain
\begin{align}
(VQ_{\lambda_2}V^{-1}a)_n
=
\begin{pmatrix}
\lambda_2a_{n-1}^+-\lambda_2'a_n^-\\
\lambda_2a_{n+1}^-+\lambda_2'a_n^+
\end{pmatrix}
=(S_{\lambda_2}a)_n.
\end{align}
Consequently,
\begin{align}
V\mathscr W_\varkappa V^{-1}
=Q_{\lambda_1,-\alpha,-\xi}S_{\lambda_2}.
\end{align}
Define
\begin{align}
\mathcal U=S_{\lambda_2}V=S_{\lambda_2}K_F\mathcal F.
\end{align}
Multiplication on the left by $S_{\lambda_2}$ restores the
factor order in the dual walk:
\begin{align}
\begin{aligned}
\mathcal U\mathscr W_\varkappa
&=S_{\lambda_2}Q_{\lambda_1,-\alpha,-\xi}
  S_{\lambda_2}V\\
&=W_{\lambda_2,\lambda_1,-\alpha,-\xi}\mathcal U.
\end{aligned}
\end{align}
These computations may first be performed on trigonometric
polynomials. Since all the operators involved are bounded
(in fact, unitary), the identity extends to all of
$L^2(\mathbb T;\mathbb C^2)$.

In particular, for any $f\in L^2(\mathbb T;\mathbb C^2)$,
\begin{align}
(\mathscr W_\varkappa-z)f=r
\quad\Longrightarrow\quad
(W_{\lambda_2,\lambda_1,-\alpha,-\xi}-z)\mathcal Uf
=\mathcal Ur,
\qquad
\|\mathcal Ur\|_{\ell^2}=\|r\|_{L^2}.
\end{align}
Thus Fourier duality transports inhomogeneous residuals
as well as eigenvectors.

\section{Interior seed gaps by finite perturbation}\label{app:interior-seeds}
We give an interior counterpart of Proposition~\ref{prop:odd-gap-seeds}.
The proof follows the same three steps: construct a finite Fourier
profile with a nonzero resonant residual, convert that residual into
a constant symmetric splitting, and apply
Lemma~\ref{lem:symmetric-seed-splitting}.
In the interior, both components occur at each Fourier index; the
additional calculation below shows that the resonant coefficient
is nonzero for every positive winding.

\begin{proposition}[$LE$]\label{prop:appendix-interior-seeds}
Fix an irrational $\alpha$, $0<\ell<1$, and $m\in\mathbb Z$.
There exists $\varepsilon_{\alpha,\ell,m}\in(0,\ell)$ such that,
whenever $0<\lambda<\varepsilon_{\alpha,\ell,m}$, each value
$\rho_*$ satisfying $2\rho_*\equiv m\alpha\pmod1$ occurs on an
open gap of $\Sigma_{\ell,\lambda,\alpha}$ and of
$\Sigma_{\lambda,\ell,\alpha}$.
\end{proposition}

For the perturbative construction, fix $m>0$ and one permitted
value $\rho_*$. Set
\begin{align}
 q=e^{2\pi i\alpha},\qquad \theta=2\pi\rho_*\pmod{2\pi},
 \qquad w=e^{i\theta},\qquad
 \sigma=w e^{-\pi im\alpha}\in\{1,-1\}.
\end{align}
Thus $w^2=q^m$ and $\sin\theta\ne0$.
Write $A_{t,\lambda}$ for \eqref{eq:real-cocycle} at couplings
$(\ell,\lambda)$, allowing $\lambda=0$, and let
$c=\sqrt{1-\ell^2}$ and
$p_\lambda(x)=\sqrt{1-\lambda^2\sin^2(2\pi x)}$.
Choose $t_0$ so that
\begin{align}
 \cos t_0=\ell\cos\theta,\qquad
 \operatorname{sgn}\sin t_0=\operatorname{sgn}\sin\theta.
\end{align}
Near $t_0$, there are real analytic $\vartheta(t)$ and
$S(t)\in\mathrm{SL}(2,\mathbb R)$ such that
\begin{align}
 S(t)^{-1}A_{t,0}S(t)=R_{\vartheta(t)},\qquad
 \vartheta(t_0)=\theta,\qquad
 \vartheta'(t_0)=\frac{\sin t_0}{\ell\sin\theta}>0.
\end{align}
These follow by diagonalizing the elliptic constant matrix $A_{t,0}$;
the choice of sign fixes the orientation of its rotation.
With $Q(t)=U_CS(t)U_C^{-1}$, define
\begin{align}\label{eq:appendix-interior-diagonal-gauge}
 \widehat M_{t,\lambda}(x)
 :=Q(t)^{-1}p_\lambda(x)\mathcal S_{e^{it}}(x)Q(t)
 =\begin{pmatrix}e^{i\vartheta(t)}&0\\0&e^{-i\vartheta(t)}\end{pmatrix}
   +\lambda\sin(2\pi x)V(t).
\end{align}
Here $V(t)$ is the conjugate by $Q(t)$ of
$\ell^{-1}\left(\begin{smallmatrix}c&-1\\-1&c\end{smallmatrix}\right)$.
It has trace $2c/\ell$ and determinant $-1$.
Moreover, $\det\widehat M_{t,\lambda}=1-\lambda^2\sin^2(2\pi x)$,
so the coefficient linear in $\lambda\sin(2\pi x)$ vanishes.
Together with the conjugate symmetry of these matrices, this gives,
after a constant rotation of $S(t)$ if necessary,
\begin{align}\label{eq:appendix-interior-perturbation-matrix}
 V:=V(t_0)=
 \begin{pmatrix}-i\mu w&\gamma\\\gamma&i\mu w^{-1}\end{pmatrix},
 \qquad \mu=\frac{c}{\ell\sin\theta}\ne0,
 \qquad \gamma=\sqrt{1+\mu^2}>0.
\end{align}
This rotation commutes with $R_{\vartheta(t)}$ and preserves the
diagonalization above.

\begin{lemma}[$L^*$]
\label{lem:appendix-interior-resonant-profile}
For sufficiently small real $\lambda$, there are real-valued
analytic functions $t(\lambda),s(\lambda)$ and a vector-valued
trigonometric polynomial $v=v_\lambda$, whose coefficients are real
analytic in $\lambda$, such that
\begin{align}
 t=t_0+O(\lambda^2),\qquad s=1+O(\lambda^2),\qquad
 v=e_1+O(\lambda),
\end{align}
and
\begin{align}\label{eq:appendix-interior-resonant-residual}
 \widehat M_{t(\lambda),\lambda}v-ws(\lambda)v(\cdot+\alpha)
 =c_m\lambda^m e^{-2\pi imx}e_2+O(\lambda^{m+1}),
 \qquad c_m\ne0.
\end{align}
The errors are uniform in $x$.
\end{lemma}

\begin{proof}
\emph{Finite cancellation.}
As in Lemma~\ref{lem:boundary-resonant-profile}, introduce a real
amplitude parameter $s$ and cancel all retained Fourier coefficients
except the second resonance. This time set
\begin{align}
 v(x)=\sum_{k=-m}^m a_k e^{2\pi ikx},\qquad a_k\in\mathbb C^2,
 \qquad a_{0,1}=1,\quad a_{-m,2}=0,
\end{align}
and set $a_k=0$ outside this interval. With $\kappa=1/(2i)$, the
coefficient of the residual at index $k$ is
\begin{align}
 F_k=
 \left[\begin{pmatrix}e^{i\vartheta(t)}&0\\0&e^{-i\vartheta(t)}\end{pmatrix}
       -wsq^k I\right]a_k
 +\kappa\lambda V(t)(a_{k-1}-a_{k+1}).
\end{align}
Solve $F_{k,j}=0$ for $-m\le k\le m$ and $j=1,2$, except for
$(k,j)=(-m,2)$, using $t,s$ and the unprescribed components of $a_k$.
At $\lambda=0$, the solution is $t=t_0$, $s=1$, $a_0=e_1$,
and $a_k=0$ for $k\ne0$. The diagonal coefficient at index $k$ is
\begin{align}
 \operatorname{diag}\bigl(w(1-q^k),\,w^{-1}(1-q^{m+k})\bigr).
\end{align}
Its only zero entries are $(0,1)$ and $(-m,2)$.
The derivative of the remaining resonant equation is
\begin{align}
 \delta F_{0,1}=w\bigl(i\vartheta'(t_0)\delta t-\delta s\bigr),
\end{align}
which is invertible over $\mathbb R$. All other equations have
nonzero diagonal derivatives in their respective complex unknowns.
The real analytic implicit function theorem therefore gives the
claimed solution. Its first derivative at $\lambda=0$ shows
$t-t_0,s-1=O(\lambda^2)$.

The same nearest-neighbor induction as in
Lemma~\ref{lem:boundary-resonant-profile} gives
$a_k=O(\lambda^{|k|})$. In this induction, the second component at
$k=-m$ is fixed to zero, and its first component has a nonzero
denominator. Thus the only residuals besides $F_{-m,2}$ are the
two omitted endpoints $k=\pm(m+1)$, both of order
$O(\lambda^{m+1})$.
For $1\le j\le m-1$, let
\begin{align}
 D_j=\operatorname{diag}\bigl(w(1-q^{-j}),\,
                              w^{-1}(1-q^{m-j})\bigr).
\end{align}
Successive substitution along the negative indices gives
\begin{align}\label{eq:appendix-interior-leading-coefficient}
 c_m=-\kappa^m K_m,\qquad
 K_m=e_2^T V D_{m-1}^{-1}V\cdots D_1^{-1}V e_1.
\end{align}
For $m=1$, this means $K_1=e_2^TVe_1=\gamma$.

\smallskip\noindent\emph{The surviving coefficient is nonzero.}
We verify $K_m\ne0$ by a finite coefficient calculation.
Write $\xi=q^{-1}$ and $T=\operatorname{diag}(1,w)$.
By \eqref{eq:appendix-interior-perturbation-matrix},
\begin{align}
 T^{-1}D_j^{-1}VT
 =\begin{pmatrix}(1-\xi^j)^{-1}&0\\
                  0&(1-\xi^{j-m})^{-1}\end{pmatrix}
   \begin{pmatrix}-i\mu&\gamma\\\gamma&i\mu\end{pmatrix}.
\end{align}
Consequently, starting with $u_0=1$, $v_0=0$, the recurrences
\begin{align}\label{eq:appendix-interior-coefficient-recurrence}
 (1-\xi^j)u_j=-i\mu u_{j-1}+\gamma v_{j-1},\qquad
 (1-\xi^{j-m})v_j=\gamma u_{j-1}+i\mu v_{j-1}
\end{align}
are uniquely solvable through $j=m-1$, and
$K_m=\gamma u_{m-1}+i\mu v_{m-1}$.
If $K_m=0$, they also extend through $j=m$: the first equation
determines $u_m$, while the second allows any $v_m$.
All later denominators are nonzero, so they extend to formal power
series. We use these series only to organize a calculation through
order $m+1$; no convergence is needed.

Set $f(y)=\sum_{j\ge0}u_jy^j$. Eliminating the series with coefficients
$v_j$ from \eqref{eq:appendix-interior-coefficient-recurrence}, and
using $\gamma^2=1+\mu^2$, gives
\begin{align}
 f(\xi^2y)
 -\bigl[1+\xi^{m+1}+i\mu\xi(1-\xi^m)y\bigr]f(\xi y)
 +\xi^{m+1}(1-y^2)f(y)=0.
\end{align}
Let $E(0)=1$ be the formal series determined by
$E(\xi y)=(1-y)E(y)$, and write $f=Eg$,
$g(y)=\sum_{n\ge0}g_ny^n$. Its coefficients satisfy
\begin{align}
 L_ng_n+G_ng_{n-1}=0\quad(n\ge1),\qquad g_0=1,
\end{align}
where
\begin{align}
 L_n=(1-\xi^n)(\xi^{m+1}-\xi^n),\qquad
 G_n=\xi^{m+1}-\xi^{2n-1}-i\mu\xi^n(1-\xi^m).
\end{align}
Here $L_n\ne0$ for $1\le n\le m$, while $L_{m+1}=0$.
On the other hand,
\begin{align}
 \frac{G_n}{\xi^n(1-\xi^m)}
 =\frac{\xi^{m+1-n}-\xi^{n-1}}{1-\xi^m}-i\mu
 =-\frac{\sin((m+2-2n)\pi\alpha)}{\sin(m\pi\alpha)}-i\mu
 \ne0,
\end{align}
since $\mu$ is real and nonzero. It follows successively that
$g_1,\ldots,g_m$ are nonzero. The equation at $n=m+1$ would then
give $G_{m+1}g_m=0$, a contradiction. Hence $K_m\ne0$, proving
\eqref{eq:appendix-interior-resonant-residual}.
\end{proof}

\begin{lemma}[$L^*$]
\label{lem:appendix-interior-resonant-splitting}
For sufficiently small positive $\lambda$, there are
$t(\lambda)=t_0+O(\lambda^2)$, a real analytic determinant-one
projective frame $B_\lambda=S(t_0)R_{\pi mx}+O(\lambda)$ of degree
$m$, and a nonzero constant real symmetric traceless matrix $H$ such
that
\begin{align}\label{eq:appendix-interior-resonant-splitting}
 \sigma B_\lambda(x+\alpha)^{-1}
 A_{t(\lambda),\lambda}(x)B_\lambda(x)
 =I+\lambda^mH+O(\lambda^{m+1}).
\end{align}
\end{lemma}

\begin{proof}
Apply the construction of
Lemma~\ref{lem:boundary-resonant-splitting} to the profile in
Lemma~\ref{lem:appendix-interior-resonant-profile}.
Explicitly, with $v=v_\lambda$, set
\begin{align}
 C(x)=\begin{pmatrix}
 e^{\pi imx}v_1(x)&e^{-\pi imx}\overline{v_2(x)}\\
 e^{\pi imx}v_2(x)&e^{-\pi imx}\overline{v_1(x)}
 \end{pmatrix},\qquad
 \Delta=\det C=|v_1|^2-|v_2|^2=1+O(\lambda)>0.
\end{align}
The conjugate symmetry of $\widehat M$ supplies the second column
from the first. Therefore, with $\zeta=c_me^{\pi im\alpha}$,
\begin{align}
 C(x+\alpha)^{-1}\widehat M_{t(\lambda),\lambda}(x)C(x)
 =\sigma sI+\lambda^m
   \begin{pmatrix}0&\overline\zeta\\\zeta&0\end{pmatrix}
   +O(\lambda^{m+1}).
\end{align}
As in that lemma, determinant normalization removes $s$ to the
required order. More precisely, take
\begin{align}
 B_\lambda=S(t(\lambda))\Delta^{-1/2}U_C^{-1}CU_C,
 \qquad
 H=\sigma U_C^{-1}
   \begin{pmatrix}0&\overline\zeta\\\zeta&0\end{pmatrix}U_C.
\end{align}
The scalar factor in the conjugated cocycle is
$\omega=\sqrt{\Delta(x+\alpha)/\Delta(x)}/p_\lambda(x)>0$.
Since the cocycle has determinant one, $\omega s=1+O(\lambda^{m+1})$.
This proves \eqref{eq:appendix-interior-resonant-splitting}.
The matrix $H$ is real, symmetric, and traceless, with eigenvalues
$\pm|c_m|$, so it is nonzero. Finally,
$C(x+1)=(-1)^mC(x)$ and
$B_\lambda=S(t_0)R_{\pi mx}+O(\lambda)$, giving projective degree $m$.
\end{proof}

\begin{proof}[Proof of Proposition~\ref{prop:appendix-interior-seeds}]
For $m>0$ and either permitted value $\rho_*$,
Lemma~\ref{lem:appendix-interior-resonant-splitting} gives
$D_\lambda=\lambda^mH$ with $H\ne0$ and remainder
$O(\lambda^{m+1})=o(\|D_\lambda\|)$.
Lemma~\ref{lem:symmetric-seed-splitting} then gives uniform
hyperbolicity and
\begin{align}
 \rho\bigl(\alpha,A_{t(\lambda),\lambda}\bigr)
 =\frac{m\alpha}{2}+\frac{1-\sigma}{4}
 =\rho_*\pmod1.
\end{align}
The spectral criterion in Section~\ref{Sec:one} and openness of
uniform hyperbolicity give an open spectral gap with this value.
The identity $A_{-t,\lambda}=H_0A_{t,\lambda}H_0^{-1}$, where
$H_0=\operatorname{diag}(1,-1)$, reverses rotation number and
supplies the negative windings, just as in
Proposition~\ref{prop:odd-gap-seeds}.

For $m=0$, at $\lambda=0$ and $t=0,\pi$ the cocycles are
$\pm\operatorname{diag}(a,a^{-1})$, where
$a=\ell/(1+\sqrt{1-\ell^2})<1$.
Their uniform hyperbolicity and labels $0,1/2$ persist for small
$\lambda$, giving open gaps at both values.
Taking the smaller threshold for the two choices of $\rho_*$ and
using coupling-exchange duality proves the proposition.
\end{proof}

\begin{remark}
The distinction from the boundary calculation is visible in
\eqref{eq:appendix-interior-perturbation-matrix}.
For $\ell<1$, the nonzero diagonal parameter $\mu$ prevents the
resonant coefficient from vanishing for any $m>0$.
At $\ell=1$, $\mu=0$, and each Fourier shift switches components,
recovering the parity restriction in
Lemma~\ref{lem:boundary-resonant-profile}.
In the coefficient calculation above, the factor $G_{(m+2)/2}$
then vanishes for even $m$, whereas all relevant factors remain
nonzero for odd $m$.
Thus both seed arguments use the same residual-to-splitting
construction and the same final hyperbolicity criterion.
\end{remark}

\begin{remark}
At fixed larger coupling, Proposition~\ref{prop:appendix-interior-seeds}
provides a seed below any prescribed target for each nonzero winding
and either of its two gap values. These gaps avoid $1,-1$ and hence
correspond to internal real gaps under \eqref{eq:harper-spectrum}.
Proposition~\ref{prop:gap-persistence} continues them to the target,
including equal couplings. For zero winding, the identity
\eqref{eq:diagonal-log-mean} remains valid for every $0<r\le a<1$:
its cancellation uses only $b(x+\tfrac12)=-b(x)$.
The same diagonal-cocycle argument therefore supplies the gaps
through $1,-1$ throughout the interior. This is the seed input used in Section~\ref{subsec:proof-interior}
for part~(i) of Theorem~\ref{thm:classification}.
\end{remark}

\section{Lean statement and verification index}\label{app:lean-audit}
For irrational $\alpha$, $0<\lambda_1,\lambda_2\leq1$, and $m\in\mathbb Z$,
Theorem~\ref{thm:classification} states that a permitted value
$\rho_*\in\mathbb R/\mathbb Z$, with $2\rho_*\equiv m\alpha\pmod1$,
labels an open gap of
$\Sigma_{\lambda_1,\lambda_2,\alpha}$ if and only if
\begin{align}
 \max\{\lambda_1,\lambda_2\}&<1\quad\text{or}\quad m\in2\mathbb Z+1.
\end{align}
Thus both values are open in the interior, including the critical line;
on the boundary, including the corner, exactly the odd windings are open.

Following the presentation in \cite[Appendix A]{BorgniaFaust2026}, we give
all theorem-specific definitions before the complete Lean statement.
The nine definitions and abbreviations below use only Mathlib; the
listing changes only line breaks and omits import and namespace boilerplate.
The phase $x$ is measured in turns, while the spectral angle $t$ is in radians. The LEAN code is available at \href{https://mattfaust.github.io/Dry10/uamo\_LEAN.zip}{mattfaust.github.io/Dry10/uamo\_LEAN.zip}.

\subsection{Shared operator vocabulary and Mathlib notation}\label{app:lean-common}
Both the AMO and the UAMO use the complex square-summable sequence
space. Its Lean abbreviation is
\par\nobreak\smallskip\noindent\begin{minipage}{\linewidth}\footnotesize\setlength{\parskip}{0pt}
\noindent\hspace*{0em}\texttt{noncomputable section}\par
\noindent\hspace*{0em}\texttt{open MeasureTheory Set}\par
\vspace{3pt}
\noindent\hspace*{0em}\texttt{abbrev SequenceSpace := $\ell^2$($\mathbb Z$, $\mathbb C$)}\par
\end{minipage}\par\smallskip

The bounded-operator language is also shared. The following Mathlib
notation will be used throughout the statement; it introduces no
model-specific assumptions.
\par\smallskip
\begingroup
\setlength{\tabcolsep}{4pt}\renewcommand{\arraystretch}{1.15}
\noindent\begin{tabular}{@{}p{.34\linewidth}p{.62\linewidth}@{}}
\textbf{Lean notation}&\textbf{Mathematical meaning}\\\hline
$E\to L[\mathbb C]E$&A bounded complex-linear operator on $E$.\\
\texttt{spectrum} $\mathbb C$ $A$&The spectrum of the operator $A$.\\
\texttt{Measure} $\mathbb C$&A measure on $\mathbb C$; \texttt{IsProbabilityMeasure} requires total mass one.\\
\texttt{AddCircle} $(1:\mathbb R)$&The additive quotient $\mathbb R/\mathbb Z$, used for gap values.\\
\texttt{Ioc} $(0:\mathbb R)$ \texttt{1}&The coupling interval $(0,1]$.\\
\texttt{Irrational} $\alpha$; \texttt{Odd m}&The frequency is irrational; the integer $m$ is odd.
\end{tabular}\par
\endgroup
\smallskip
The UAMO-specific additions are the two spin components and the
unit-circle spectral conventions in the next two sections.

\subsection{The UAMO family}\label{app:lean-uamo}
\noindent\emph{Two spin components.}
\texttt{Fin 2} indexes the spins. A walk is a $2\times2$ matrix of
bounded operators on \texttt{SequenceSpace}; matrix multiplication
composes the operator blocks.
\par\nobreak\smallskip\noindent\begin{minipage}{\linewidth}\footnotesize\setlength{\parskip}{0pt}
\noindent\hspace*{0em}\texttt{abbrev WalkOperator :=}\par
\noindent\hspace*{0.68em}\texttt{Matrix (Fin 2) (Fin 2) (SequenceSpace $\to$L[$\mathbb C$] SequenceSpace)}\par
\end{minipage}\par\smallskip

\noindent\emph{The local coin.}
\texttt{complement lam} is $\sqrt{1-\mathrm{lam}^2}$.
The two entries of \texttt{coin} give the action of $Q$ on a vector
$\psi$ at site $n$. Here the phase $x$ is measured in turns, so
\texttt{angle} is measured in radians.
\par\nobreak\smallskip\noindent\begin{minipage}{\linewidth}\footnotesize\setlength{\parskip}{0pt}
\noindent\hspace*{0em}\texttt{def complement (lam : $\mathbb R$) : $\mathbb R$ := Real.sqrt (1-lam\textasciicircum{}2)}\par
\vspace{3pt}
\noindent\hspace*{0em}\texttt{def coin ($\alpha$ lam x : $\mathbb R$) ($\psi$ : Fin 2 $\to$ SequenceSpace)}\par
\noindent\hspace*{1.36em}\texttt{(n : $\mathbb Z$) : Fin 2 $\to$ $\mathbb C$ :=}\par
\noindent\hspace*{0.68em}\texttt{let angle := 2*Real.pi*(x+(n:$\mathbb R$)*$\alpha$)}\par
\noindent\hspace*{0.68em}\texttt{![((lam*Real.cos angle:$\mathbb R$):$\mathbb C$)*$\psi$ 0 n + Complex.I*(complement lam:$\mathbb C$)*$\psi$ 0 n -}\par
\noindent\hspace*{2.04em}\texttt{((lam*Real.sin angle:$\mathbb R$):$\mathbb C$)*$\psi$ 1 n,}\par
\noindent\hspace*{1.36em}\texttt{((lam*Real.sin angle:$\mathbb R$):$\mathbb C$)*$\psi$ 0 n + ((lam*Real.cos angle:$\mathbb R$):$\mathbb C$)*$\psi$ 1 n -}\par
\noindent\hspace*{2.04em}\texttt{Complex.I*(complement lam:$\mathbb C$)*$\psi$ 1 n]}\par
\end{minipage}\par\smallskip

\noindent\emph{The physical walk.}
For a fixed frequency $\alpha$, the expression
\texttt{W lam$_1$ lam$_2$ x} means the walk at the two couplings and
phase $x$. The predicate \texttt{IsUAMO} specifies $W=SQ$ by its action
on every vector at every site. The allowed coupling ranges are stated
in the theorem.
\par\nobreak\smallskip\noindent\begin{minipage}{\linewidth}\footnotesize\setlength{\parskip}{0pt}
\noindent\hspace*{0em}\texttt{def IsUAMO ($\alpha$ : $\mathbb R$) (W : $\mathbb R$ $\to$ $\mathbb R$ $\to$ $\mathbb R$ $\to$ WalkOperator) : Prop :=}\par
\noindent\hspace*{0.68em}\texttt{$\forall$ lam$_1$ lam$_2$ x ($\psi$ : Fin 2 $\to$ SequenceSpace) (n : $\mathbb Z$),}\par
\noindent\hspace*{1.36em}\texttt{($\sum$ j : Fin 2, W lam$_1$ lam$_2$ x 0 j ($\psi$ j) n) =}\par
\noindent\hspace*{2.04em}\texttt{(lam$_1$:$\mathbb C$)*coin $\alpha$ lam$_2$ x $\psi$ (n-1) 0 -}\par
\noindent\hspace*{2.72em}\texttt{(complement lam$_1$:$\mathbb C$)*coin $\alpha$ lam$_2$ x $\psi$ n 1 $\land$}\par
\noindent\hspace*{1.36em}\texttt{($\sum$ j : Fin 2, W lam$_1$ lam$_2$ x 1 j ($\psi$ j) n) =}\par
\noindent\hspace*{2.04em}\texttt{(lam$_1$:$\mathbb C$)*coin $\alpha$ lam$_2$ x $\psi$ (n+1) 1 +}\par
\noindent\hspace*{2.72em}\texttt{(complement lam$_1$:$\mathbb C$)*coin $\alpha$ lam$_2$ x $\psi$ n 0}\par
\end{minipage}\par\smallskip

\subsection{Unit-circle DOS and gap labels}\label{app:lean-circle}
The spectral values of the UAMO lie on the unit circle. The following
definitions specify the DOS and labels for that geometry.

\noindent\emph{Density of states.}
\texttt{IsDOS} requires a probability measure supported on the unit
circle. Its $k$th moment is the phase average of the origin matrix
elements of $W^k$, with equal weight for the two spins. The vector
\texttt{lp.single 2 0 1} is the unit sequence at the origin;
$\forall^{\mathrm m}z\,\partial\nu$ means ``for $\nu$-almost every $z$''.
\par\nobreak\smallskip\noindent\begin{minipage}{\linewidth}\footnotesize\setlength{\parskip}{0pt}
\noindent\hspace*{0em}\texttt{def IsDOS (W : $\mathbb R$ $\to$ WalkOperator) ($\nu$ : Measure $\mathbb C$) : Prop :=}\par
\noindent\hspace*{0.68em}\texttt{IsProbabilityMeasure $\nu$ $\land$ ($\forall^{\mathrm{m}}$ z $\partial$$\nu$, $\Vert$z$\Vert$=1) $\land$}\par
\noindent\hspace*{0.68em}\texttt{$\forall$ k : $\mathbb N$, ($\int$ z, z\textasciicircum{}k $\partial$$\nu$) =}\par
\noindent\hspace*{1.36em}\texttt{$\int$ x in (0:$\mathbb R$)..1,}\par
\noindent\hspace*{2.04em}\texttt{((((W x)\textasciicircum{}k) 0 0 (lp.single 2 0 (1:$\mathbb C$))) 0 +}\par
\noindent\hspace*{2.38em}\texttt{(((W x)\textasciicircum{}k) 1 1 (lp.single 2 0 (1:$\mathbb C$))) 0)/2}\par
\end{minipage}\par\smallskip

\noindent\emph{Reading the circular DOS.}
\texttt{arc t} fixes the cut at the point $e^{it}$, with $t$ in radians.
The displayed inequalities specify the endpoint convention exactly;
only the intersection with the unit circle carries DOS mass.
\par\nobreak\smallskip\noindent\begin{minipage}{\linewidth}\footnotesize\setlength{\parskip}{0pt}
\noindent\hspace*{0em}\texttt{def arc (t : $\mathbb R$) : Set $\mathbb C$ :=}\par
\noindent\hspace*{0.68em}\texttt{let z := Complex.exp ((t:$\mathbb C$)*Complex.I)}\par
\noindent\hspace*{0.68em}\texttt{if 0<z.im then \{w | 0<w.im $\land$ z.re$\le$w.re\}}\par
\noindent\hspace*{0.68em}\texttt{else \{w | 0<w.im $\lor$ (w.im<0 $\land$ w.re<z.re)\}}\par
\end{minipage}\par\smallskip

\noindent\emph{An open labelled gap.}
\texttt{OpenGap W} $\nu\,\rho$ means that one angular interval of
positive radius lies outside the spectrum at every phase. On the
whole interval, the mass of \texttt{arc s} under $\nu$, read modulo one,
is $\rho$. The chosen interval can be any open subinterval of the gap;
its endpoints need not be spectral edges. In Lean, \texttt{$\nu$.real}
reads the finite measure as a real number.
\par\nobreak\smallskip\noindent\begin{minipage}{\linewidth}\footnotesize\setlength{\parskip}{0pt}
\noindent\hspace*{0em}\texttt{def OpenGap (W : $\mathbb R$ $\to$ WalkOperator) ($\nu$ : Measure $\mathbb C$)}\par
\noindent\hspace*{1.36em}\texttt{($\rho$ : AddCircle (1:$\mathbb R$)) : Prop :=}\par
\noindent\hspace*{0.68em}\texttt{$\exists$ t $\delta$ : $\mathbb R$, 0<$\delta$ $\land$ $\forall$ s : $\mathbb R$, |s-t|<$\delta$ $\to$}\par
\noindent\hspace*{1.36em}\texttt{($\forall$ x : $\mathbb R$, Complex.exp ((s:$\mathbb C$)*Complex.I) $\notin$ spectrum $\mathbb C$ (W x)) $\land$}\par
\noindent\hspace*{1.36em}\texttt{($\nu$.real (arc s) : AddCircle (1:$\mathbb R$))=$\rho$}\par
\end{minipage}\par\smallskip

\noindent\emph{One integer winding label.}
\texttt{HasWinding} means $2\rho=m\alpha\pmod1$; the symbol $\bullet$
denotes repeated addition. The same signed integer $m$ labels both
solutions for $\rho$. The theorem tests both values, without a second
integer index.
\par\nobreak\smallskip\noindent\begin{minipage}{\linewidth}\footnotesize\setlength{\parskip}{0pt}
\noindent\hspace*{0em}\texttt{def HasWinding ($\alpha$ : $\mathbb R$) (m : $\mathbb Z$) ($\rho$ : AddCircle (1:$\mathbb R$)) : Prop :=}\par
\noindent\hspace*{0.68em}\texttt{(2:$\mathbb N$) $\bullet$ $\rho$ = (((m:$\mathbb R$)*$\alpha$:$\mathbb R$) : AddCircle (1:$\mathbb R$))}\par
\end{minipage}\par\smallskip

\subsection{Classification and the complete theorem}\label{app:lean-theorem}
\noindent\emph{The classification clause.}
For one coupling pair and its DOS, \texttt{GapClassification} names
the conclusion: every permitted gap value is open exactly when both
couplings are below one or its winding integer is odd.
\par\nobreak\smallskip\noindent\begin{minipage}{\linewidth}\footnotesize\setlength{\parskip}{0pt}
\noindent\hspace*{0em}\texttt{def GapClassification ($\alpha$ lam$_1$ lam$_2$ : $\mathbb R$)}\par
\noindent\hspace*{1.36em}\texttt{(W : $\mathbb R$ $\to$ WalkOperator) ($\nu$ : Measure $\mathbb C$) : Prop :=}\par
\noindent\hspace*{0.68em}\texttt{$\forall$ (m : $\mathbb Z$) ($\rho$ : AddCircle (1:$\mathbb R$)), HasWinding $\alpha$ m $\rho$ $\to$}\par
\noindent\hspace*{1.36em}\texttt{(OpenGap W $\nu$ $\rho$ $\leftrightarrow$ (lam$_1$ < 1 $\land$ lam$_2$ < 1) $\lor$ Odd m)}\par
\end{minipage}\par\smallskip

\noindent\emph{Reading the quantifiers.}
For every irrational frequency, the theorem constructs a family $W$
satisfying \texttt{IsUAMO}. For every coupling pair in $(0,1]^2$, it
then supplies a unique measure satisfying both \texttt{IsDOS} and
\texttt{GapClassification}. Thus $\exists!\,\nu$ scopes over the entire
conjunction below. Expanding the displayed classification definition
recovers the full quantified conclusion, including both circle values,
zero and negative windings, the critical diagonal, both boundary edges,
and the corner.
\par\nobreak\smallskip\noindent\begin{minipage}{\linewidth}\footnotesize\setlength{\parskip}{0pt}
\noindent\hspace*{0em}\texttt{theorem readable\_uamo :}\par
\noindent\hspace*{1.36em}\texttt{$\forall$ $\alpha$ : $\mathbb R$, Irrational $\alpha$ $\to$}\par
\noindent\hspace*{1.36em}\texttt{$\exists$ W : $\mathbb R$ $\to$ $\mathbb R$ $\to$ $\mathbb R$ $\to$ WalkOperator, IsUAMO $\alpha$ W $\land$}\par
\noindent\hspace*{2.04em}\texttt{$\forall$ lam$_1$ $\in$ Ioc (0:$\mathbb R$) 1, $\forall$ lam$_2$ $\in$ Ioc (0:$\mathbb R$) 1,}\par
\noindent\hspace*{2.04em}\texttt{$\exists$! $\nu$ : Measure $\mathbb C$,}\par
\noindent\hspace*{2.72em}\texttt{IsDOS (W lam$_1$ lam$_2$) $\nu$ $\land$}\par
\noindent\hspace*{2.72em}\texttt{GapClassification $\alpha$ lam$_1$ lam$_2$ (W lam$_1$ lam$_2$) $\nu$ :=}\par
\noindent\hspace*{0.68em}\texttt{PaperUAMO.Submission.uamo}\par
\end{minipage}\par\smallskip

The last line is the complete proof of this presentation: Lean accepts
the original verified theorem directly. A separate check expands
\texttt{GapClassification} and recovers the original theorem type.
The shorter statement therefore has exactly the same hypotheses and
conclusion. No seed, Lyapunov or normal-form assumption is introduced.
The source files for the definitions, statement and checks are listed
with the index below.

\subsection{Result-by-result index}
The index follows the numbering of this draft: there are thirty numbered
theorems, propositions, lemmas and corollaries. The DOS correspondence
is now unnumbered supporting text and is listed separately.
$L^*$ records the checked written argument; $LE$ records a checked
conclusion with the proof replacement specified below. These are local
statuses: an $L^*$ entry does not upgrade an $LE$ dependency.
All paths are relative to the reduced Lean package.

\begingroup\footnotesize
\setlength{\tabcolsep}{4pt}\renewcommand{\arraystretch}{1.12}
\noindent\begin{tabular}{@{}p{.12\linewidth}p{.07\linewidth}p{.76\linewidth}@{}}
Result&Status&Lean files\\\hline
\end{tabular}\par\nobreak
\noindent\begin{tabular}{@{}p{.12\linewidth}p{.07\linewidth}p{.76\linewidth}@{}}\ref{thm:interior}&$LE$&\leanname{PaperUAMO/Submission/Theorem.lean}\\\end{tabular}\par
\noindent\begin{tabular}{@{}p{.12\linewidth}p{.07\linewidth}p{.76\linewidth}@{}}\ref{thm:classification}&$L^*$&\leanname{PaperUAMO/Submission/Theorem.lean}\\\end{tabular}\par
\noindent\begin{tabular}{@{}p{.12\linewidth}p{.07\linewidth}p{.76\linewidth}@{}}\ref{lem:opening}&$L^*$&\leanname{UAMODraft/IntervalOpening.lean}\\\end{tabular}\par
\noindent\begin{tabular}{@{}p{.12\linewidth}p{.07\linewidth}p{.76\linewidth}@{}}\ref{lem:gauge}&$L^*$&\leanname{UAMODraft/GeneralScalarGauge.lean}\\\end{tabular}\par
\noindent\begin{tabular}{@{}p{.12\linewidth}p{.07\linewidth}p{.76\linewidth}@{}}\ref{lem:wronskian}&$L^*$&\leanname{UAMODraft/WronskianEuclidean.lean}\\\end{tabular}\par
\noindent\begin{tabular}{@{}p{.12\linewidth}p{.07\linewidth}p{.76\linewidth}@{}}\ref{prop:boundary-even-closure}&$L^*$&\leanname{UAMO/BoundaryStableParity.lean}; \leanname{UAMO/PhysicalBoundaryParity.lean}\\\end{tabular}\par
\noindent\begin{tabular}{@{}p{.12\linewidth}p{.07\linewidth}p{.76\linewidth}@{}}\ref{prop:odd-gap-seeds}&$L^*$&\leanname{UAMODraft/LiteralBoundarySeeds.lean}; \leanname{UAMODraft/AnalyticSeedLabels.lean}\\\end{tabular}\par
\noindent\begin{tabular}{@{}p{.12\linewidth}p{.07\linewidth}p{.76\linewidth}@{}}\ref{lem:boundary-resonant-profile}&$L^*$&\leanname{UAMODraft/UniformResidual.lean}; \leanname{UAMODraft/PhysicalProfile.lean}; \leanname{UAMODraft/AnalyticBranch.lean}; \leanname{UAMODraft/Center.lean}; \leanname{UAMODraft/CoefficientFormula.lean}\\\end{tabular}\par
\noindent\begin{tabular}{@{}p{.12\linewidth}p{.07\linewidth}p{.76\linewidth}@{}}\ref{lem:boundary-resonant-splitting}&$L^*$&\leanname{PaperUAMO/BoundarySplittingTheorem.lean}\\\end{tabular}\par
\noindent\begin{tabular}{@{}p{.12\linewidth}p{.07\linewidth}p{.76\linewidth}@{}}\ref{lem:symmetric-seed-splitting}&$L^*$&\leanname{PaperUAMO/SymmetricSplittingCircle.lean}\\\end{tabular}\par
\noindent\begin{tabular}{@{}p{.12\linewidth}p{.07\linewidth}p{.76\linewidth}@{}}\ref{lem:reflection}&$L^*$&\leanname{UAMODraft/ScalarBlocks.lean}; \leanname{UAMODraft/ScalarBlocksSymmetry.lean}; \leanname{UAMODraft/ScalarBlocksDOSSymmetry.lean}\\\end{tabular}\par
\noindent\begin{tabular}{@{}p{.12\linewidth}p{.07\linewidth}p{.76\linewidth}@{}}\ref{lem:cosine}&$L^*$&\leanname{UAMODraft/ScalarBlocksSpectrum.lean}\\\end{tabular}\par
\noindent\begin{tabular}{@{}p{.12\linewidth}p{.07\linewidth}p{.76\linewidth}@{}}\ref{lem:block-symmetry}&$L^*$&\leanname{UAMODraft/ScalarBlocksHarper.lean}; \leanname{UAMODraft/DraftGaugeHarper.lean}; \leanname{UAMODraft/DraftGaugeSpectrum.lean}\\\end{tabular}\par
\noindent\begin{tabular}{@{}p{.12\linewidth}p{.07\linewidth}p{.76\linewidth}@{}}\ref{lem:height-label-properties}&$L^*$&\leanname{UAMODraft/OffSpectrumHeights.lean}\\\end{tabular}\par
\noindent\begin{tabular}{@{}p{.12\linewidth}p{.07\linewidth}p{.76\linewidth}@{}}\ref{lem:gap-label-constancy}&$LE$&\leanname{UAMO/GapProjectionTrace.lean}\\\end{tabular}\par
\noindent\begin{tabular}{@{}p{.12\linewidth}p{.07\linewidth}p{.76\linewidth}@{}}\ref{lem:height-correspondence}&$LE$&\leanname{UAMO/CMVThoulessClosed.lean}; \leanname{UAMO/ScalarThouless.lean}\\\end{tabular}\par
\noindent\begin{tabular}{@{}p{.12\linewidth}p{.07\linewidth}p{.76\linewidth}@{}}\ref{cor:spectral-baseline}&$LE$&\leanname{UAMO/ScalarSpectralBaseline.lean}\\\end{tabular}\par
\noindent\begin{tabular}{@{}p{.12\linewidth}p{.07\linewidth}p{.76\linewidth}@{}}\ref{lem:height-usc}&$L^*$&\leanname{UAMODraft/TotalHeightUSC.lean}\\\end{tabular}\par
\noindent\begin{tabular}{@{}p{.12\linewidth}p{.07\linewidth}p{.76\linewidth}@{}}\ref{lem:gap-maximum}&$L^*$&\leanname{UAMODraft/GapHeightMaximumByIntegration.lean}; \leanname{UAMODraft/GapHeightBranches.lean}\\\end{tabular}\par
\noindent\begin{tabular}{@{}p{.12\linewidth}p{.07\linewidth}p{.76\linewidth}@{}}\ref{lem:derivative-bound}&$L^*$&\leanname{UAMODraft/GapHeightPairing.lean}\\\end{tabular}\par
\noindent\begin{tabular}{@{}p{.12\linewidth}p{.07\linewidth}p{.76\linewidth}@{}}\ref{thm:height-monotonicity}&$L^*$&\leanname{UAMODraft/GapHeightComplete.lean}; \leanname{UAMO/MagneticResponse.lean}\\\end{tabular}\par
\noindent\begin{tabular}{@{}p{.12\linewidth}p{.07\linewidth}p{.76\linewidth}@{}}\ref{lem:radial-cone}&$L^*$&\leanname{UAMODraft/MixedDerivativeOrder.lean}; \leanname{UAMODraft/TwistDerivative.lean}; \leanname{UAMODraft/Coordinates.lean}\\\end{tabular}\par
\noindent\begin{tabular}{@{}p{.12\linewidth}p{.07\linewidth}p{.76\linewidth}@{}}\ref{lem:walk-stationarity}&$L^*$&\leanname{UAMO/MagneticStationarity.lean}\\\end{tabular}\par
\noindent\begin{tabular}{@{}p{.12\linewidth}p{.07\linewidth}p{.76\linewidth}@{}}\ref{lem:axis-terms}&$L^*$&\leanname{UAMO/MagneticBalanceModel.lean}; \leanname{UAMO/CenteredCompression.lean}\\\end{tabular}\par
\noindent\begin{tabular}{@{}p{.12\linewidth}p{.07\linewidth}p{.76\linewidth}@{}}\ref{lem:gap-resolvent}&$L^*$&\leanname{UAMODraft/ResolventConverse.lean}\\\end{tabular}\par
\noindent\begin{tabular}{@{}p{.12\linewidth}p{.07\linewidth}p{.76\linewidth}@{}}\ref{cor:resolvent-parity}&$L^*$&\leanname{UAMO/CompressedEigenline.lean}; \leanname{UAMO/MagneticResponse.lean}; \leanname{UAMODraft/GapHeightComplete.lean}\\\end{tabular}\par
\noindent\begin{tabular}{@{}p{.12\linewidth}p{.07\linewidth}p{.76\linewidth}@{}}\ref{prop:gap-persistence}&$L^*$&\leanname{UAMODraft/GapHeightPersistence.lean}; \leanname{UAMODraft/DraftGapContinuation.lean}\\\end{tabular}\par
\noindent\begin{tabular}{@{}p{.12\linewidth}p{.07\linewidth}p{.76\linewidth}@{}}\ref{prop:appendix-interior-seeds}&$LE^{\dagger}$&\leanname{PaperUAMO/InteriorSeedPublic.lean}\\\end{tabular}\par
\noindent\begin{tabular}{@{}p{.12\linewidth}p{.07\linewidth}p{.76\linewidth}@{}}\ref{lem:appendix-interior-resonant-profile}&$L^*$&\leanname{PaperUAMO/InteriorProfileTheorem.lean}; \leanname{PaperUAMO/InteriorCoefficientIdentification.lean}\\\end{tabular}\par
\noindent\begin{tabular}{@{}p{.12\linewidth}p{.07\linewidth}p{.76\linewidth}@{}}\ref{lem:appendix-interior-resonant-splitting}&$L^*$&\leanname{PaperUAMO/InteriorSplittingTheorem.lean}\\\end{tabular}\par
\endgroup

The original public vocabulary and theorem are in
\leanname{PaperUAMO/Submission/Definitions.lean} and
\leanname{PaperUAMO/Submission/Theorem.lean}.
The readable statement is supplied under
\leanname{PaperUAMO/Submission/}:
\leanname{ReadableDefinitions.lean} names the classification clause,
\leanname{ReadableTheorem.lean} proves the displayed theorem, and
\leanname{ReadableSignature.lean} checks its fully expanded type.

The unnumbered identities in Appendix~\ref{app:duality} are $L^*$
(\leanname{UAMODraft/FourierDuality.lean}).
The DOS and semicircle correspondence in \eqref{eq:real-unitary-labels}
is $L^*$ (\leanname{UAMO/PhysicalDOSLabels.lean} and
\leanname{UAMO/ScalarIDSContinuity.lean}); the rotation--DOS identification
in \eqref{eq:dos-rotation} retains $LE$
(\leanname{UAMO/PhysicalRotationDOS.lean}).
The magnetic-inverse remark is unused ($W$).

Theorem~\ref{thm:interior} retains $LE$: its conclusion is checked through
the finite-seed and height argument, while the normal-form input in
Section~\ref{Sec:one} remains the external input of
\cite[Theorems 3.2--3.3]{AYZ24}.
Lemma~\ref{lem:gap-label-constancy} replaces the cited $K$-theory input by
nearby-projection intertwining and trace cyclicity.
Lemma~\ref{lem:height-correspondence} and
Corollary~\ref{cor:spectral-baseline} use checked internal counterparts
of the Thouless formula and the spectral Lyapunov theorem.

$\dagger$ For Proposition~\ref{prop:appendix-interior-seeds}, the
nonzero-winding construction follows the written two-component argument;
Lemmas~\ref{lem:appendix-interior-resonant-profile} and
\ref{lem:appendix-interior-resonant-splitting} are $L^*$.
For zero winding, the checked proof uses the explicit diagonal-cocycle
calculation in Section~\ref{subsec:proof-interior} and the closing remark
of Appendix~\ref{app:interior-seeds}, rather than formalizing the local
$\lambda=0$ openness sentence in this proposition's proof. The full entry
is therefore conservatively $LE$ for the present wording; the unconditional
Lean seed conclusion is unchanged. Theorem~\ref{thm:classification}'s
$L^*$ entry refers to the seed-height assembly described in Section~\ref{sec:fullproof}.
For Lemma~\ref{lem:symmetric-seed-splitting}, the formalization uses the
Euclidean operator norm and its induced uniform norm.

The proof package, including the readable statement and its fully expanded
guard, has a successful fresh build and independent kernel replay of
$566$ modules and $7{,}436$ safe logical declarations,
including $6{,}249$ theorems. Its logical axioms are
\texttt{propext}, \texttt{Classical.choice}, and \texttt{Quot.sound};
pinned Lean/Mathlib library objects are the trusted library base.
This appendix prints the full public statement in parts; it does not claim that the external Section~\ref{Sec:one}
normal-form existence proof has been formalized.

\bibliographystyle{plain}
\bibliography{reference}

\end{document}